\documentclass[acmsmall,screen]{acmart}

\usepackage[utf8]{inputenc}
\usepackage[english]{babel}

\usepackage{graphicx}
\DeclareGraphicsExtensions{.pdf,.png,.jpg}
\usepackage[ruled,vlined,linesnumbered]{algorithm2e}

\usepackage{booktabs}
\usepackage{wrapfig}
\usepackage{multirow}
\usepackage{enumitem}
\usepackage{makecell}
\usepackage{relsize}
\usepackage{placeins}

\usepackage{xcolor}
\usepackage[most]{tcolorbox}
\usepackage[normalem]{ulem}
\usepackage{colortbl}
\usepackage{booktabs} 
\usepackage{tabularx}

\definecolor{asyncbg}{RGB}{248,248,248}
\definecolor{asyncframe}{RGB}{150,150,150}
\definecolor{asynckeyword}{RGB}{120,55,135}
\definecolor{asyncapi}{RGB}{38,94,120}
\definecolor{asyncrisk}{RGB}{150,70,50}
\definecolor{codegray}{RGB}{105,105,105}
\definecolor{codepurple}{RGB}{120,55,135}

\colorlet{darkgreen}{codegray}
\colorlet{codegreen}{codegray}

\lstdefinelanguage{AsyncPython}[]{Python}{
  morekeywords=[1]{async,await,import,as,return,def,for,in,with,try,finally,except},
  morekeywords=[2]{create_task,gather,acquire,release},
  morekeywords=[3]{requests,threading,local,get},
  sensitive=true
}

\lstdefinestyle{aegisasync-python}{
  language=AsyncPython,
  basicstyle=\ttfamily\tiny,
  keywordstyle=[1]\color{asynckeyword}\bfseries,
  keywordstyle=[2]\color{asyncapi}\bfseries,
  keywordstyle=[3]\color{asyncrisk}\bfseries,
  stringstyle=\color{asyncrisk},
  commentstyle=\color{codegray},
  identifierstyle=\color{black},
  columns=fullflexible,
  keepspaces=true,
  breaklines=true,
  aboveskip=1pt,
  belowskip=0pt,
  numbers=left,
  numberstyle=\tiny\color{codegray},
  xleftmargin=1.8em,
  numbersep=0.4em,
  showstringspaces=false
}

\tcbset{
  asyncbox/.style={
    enhanced,
    colback=white,
    colframe=black!55,
    boxrule=0.35pt,
    arc=1mm,
    left=2pt,
    right=2pt,
    top=2pt,
    bottom=2pt,
    boxsep=1pt
  }
}

\usepackage{xspace}
\usepackage{tikz}
\usetikzlibrary{arrows.meta,positioning,shapes,calc} 

\newcommand{\find}[1]{%
\begin{tcolorbox}[
  tile,
  size=fbox,
  boxsep=1.3mm,
  boxrule=0pt,
  top=0.5mm,
  bottom=0.5mm,
  borderline west={1mm}{0pt}{black!45},
  colback=black!4!white
]
#1
\end{tcolorbox}}

\newboolean{COMMENTSON} 
\setboolean{COMMENTSON}{true}   
\ifthenelse{\boolean{COMMENTSON}}
{

}

\definecolor{DarkOrange}{rgb}{0.8,0.3,0.0} 
\definecolor{DarkCyan}{rgb}{0.0, 0.55, 0.55}
\definecolor{codegreen}{rgb}{0,0.6,0}
\definecolor{codegray}{rgb}{0.5,0.5,0.5}
\definecolor{codepurple}{rgb}{0.58,0,0.82}
\definecolor{backcolour}{rgb}{0.95,0.95,0.92}

\newcommand{\blackcircleone}[1]{%
  \tikz[baseline=(char.base)]{
    \node[shape=circle, fill=black, text=white, inner sep=1pt] (char) {\small #1};
  }%
}

\newcommand{\papertitle}{AegisAsync: Symbolic Async-Event Simulation for Risks in Python Asynchronous Programs}
\newcommand{\papertitleshort}{AegisAsync}

\newcommand{\tech}{\mbox{\textsc{AegisAsync}}\xspace}

\newcommand{\paperkeywords}{Python async programming, bytecode analysis, static analysis, event-loop security, vulnerability detection}

\newcommand{\appendixcontent}{%
  \appendix
}

\newcommand{\artifact}[1]{%
  \href{https://anonymous.4open.science/r/AegisAsync_Anonymous-027E}{\underline{#1}}%
}

\setcopyright{none}
\renewcommand\footnotetextcopyrightpermission[1]{}

\begin{document}

\title[\papertitleshort]{\papertitle}

\author{Baihong Chen}
\affiliation{%
  \institution{Utah State University}
  \country{USA}
}

\author{Wen Li}
\affiliation{%
  \institution{Utah State University}
  \country{USA}
}



\begin{abstract}

Security and robustness risks in Python async programs depend
on event order, object identity, and execution after exceptions
or cancellation.
Detecting these risks requires connecting related operations
across function calls and tracking their effects along execution
paths.
We present {\tech}, a bytecode-based analysis of Python async
risks for CPython~3.10--3.14.
Its Async Event Transition Graph combines bytecode control flow,
resolved calls, async events, and symbolic object relations.
Bounded path-local simulation tracks event order and object
states and explores cancellation at await suspensions.
Six detector plugins evaluate the resulting traces and states
to produce candidates with supporting witnesses.
On AsyncBench, our benchmark of 180 labeled risks, {\tech}
detects all labeled risks with zero false positives on each
of the five versions.
Across five path budgets on 2,500 PyPI package versions, source review confirms 457 of 910
{\tech} candidate records (50.2\%), representing 457 unique bugs; 
across three \textsc{flake8-async} samples, 
70 of 2,187 review judgments are confirmed (3.2\%).
By submission, 
166 PoCs reproduce {\tech} findings, 
and we have reported \textbf{157} bugs to developers, of which \textbf{16} have been confirmed and fixed.

\end{abstract}


\maketitle

\section{Introduction}\label{sec:intro}

Python asynchronous programming is widely used 
in web services, API servers, database clients, and AI-serving systems. 
These systems use features such as \texttt{await}, task creation, cancellation, 
and asynchronous resource management to support high concurrency~\cite{pep3156,pep492,python-asyncio}.
However, 
asynchronous execution can also introduce security and robustness risks. 
Blocking operations can stall the event loop. 
Unobserved tasks can hide failures. 
Input-dependent task creation and unbounded queues can exhaust resources. 
Cancellation can interrupt cleanup or be suppressed unintentionally. 
External operations without timeouts can remain pending indefinitely~\cite{python-asyncio-dev,python-asyncio-task,lu2008learning}.
Detecting these risks requires more than identifying individual API calls. 
Their effects often depend on the order of events, 
the identities and lifetimes of runtime objects, 
and how execution proceeds through exceptions and cancellation.

These risks matter across the Python ecosystem. 
Our three-year PyPI study shows that 
asynchronous programming is widespread 
across major Python application domains~\cite{pypi-index-api,pypi-json-api}. 
Public advisories also document serious async-related failures, 
including denial-of-service and resource-management errors~\cite{osv-dev,github-advisory-database,cve-program}.
Existing tools capture only part of this problem. 
General Python analyzers reason about security patterns, types, or call relations~\cite{bandit,pyre,salis2021pycg}. 
Async-aware linters such as \textsc{flake8-async}~\cite{flake8-async} 
detect known risky usage patterns. 
However, 
async risks can depend on how events 
involving the same runtime object unfold over time.
Dynamic testing can expose these failures, 
but depends on suitable inputs and execution conditions~\cite{klees2018evaluating,afl-fuzz,ossfuzz}. 
This motivates a static analysis that recovers async-event relations 
along execution paths and reports candidates with supporting evidence.

Building such an analysis raises four challenges.
\emph{(C1) Recovering async execution.}
Source code describes async behavior at a high level, 
while bytecode shows how the behavior is executed. 
The challenge is to recover this execution flow while preserving source locations for reporting.
\emph{(C2) Tracking async objects.}
Async risks can depend on how objects are created and used over time. 
The challenge is to track these objects together with the execution flow.
\emph{(C3) Tracking event order.}
Async risks can depend on the order of events along one execution path and on the same object. 
The challenge is to track these events in the order they occur.
\emph{(C4) Checking risk conditions.}
Different async risks depend on different events and object states. 
The challenge is to check these conditions and keep the evidence for each reported candidate.

To address these challenges, we present {\tech}, a static analysis
framework for finding async risks in Python bytecode.
For each target package, {\tech} analyzes the bytecode compiled for
a fixed CPython version.
The analysis begins by constructing an \emph{Async Event Transition
Graph} (AETG), which connects bytecode execution with async operations
and symbolic object relations (C1--C2).
It then explores selected execution paths over the AETG, tracking
event order, object state, and cancellation behavior (C3).
Risk-specific detectors evaluate the resulting traces and states
to identify candidates and retain the evidence supporting each
report (C4).

We evaluate {\tech} in two settings.
First, we use AsyncBench, a benchmark we construct with
180 labeled risk instances, to test evidence preservation
and detection across CPython~3.10--3.14.
{\tech} detects all 180 risks with zero false positives
on each version.
Second, we analyze 2,500 real-world Python package versions.
At the default path budget $N_p=4096$, {\tech} reports
572 candidates, of which source review confirms 259.
Deduplicating reports across all five tested path budgets
yields 910 candidates, of which 457 are confirmed (50.2\%).
In the three risk families covered by both tools, the combined
{\tech} results contain 484 candidates, compared with
5,581 diagnostics from \textsc{flake8-async} on the same packages.
Manual review across three baseline samples produces
2,187 judgments, including 70 confirmations (3.2\%)
covering 33 unique sites.
{\tech} reports candidates at all 33 sites.
By submission, 166 executable PoCs reproduce the reported
runtime behaviors across all six risk families.
This paper makes the following \textbf{contributions}:
\begin{itemize}[leftmargin=*, nosep]

  \item An \emph{Async Event Transition Graph} (AETG) that
  combines bytecode control flow, async events, and relations
  among symbolic objects.

  \item A path-local symbolic simulation that tracks event
  order and object states and explores cancellation continuations.
  Six detector plugins evaluate the resulting traces and states
  and report candidates with supporting witnesses.

  \item A study of 2,500 real-world Python package versions.
  Source review confirms 457 bugs from 910 candidates across
  five path budgets, and 166 executable PoCs reproduce the
  reported bugs.

  \item A version-aware implementation of {\tech} for
CPython~3.10--3.14, with adapters for version-specific
bytecode and a pipeline for compiling and analyzing Python packages.
\end{itemize}

\section{Background and Motivation}\label{sec:background}

\subsection{Background Knowledge}\label{sec:bg:knowledge}

\noindent
\textbf{Python asynchronous programming model.}
Python's \texttt{asyncio} framework uses cooperative scheduling:
tasks share an event loop and may suspend at an \texttt{await}
when the awaited operation cannot complete immediately
~\cite{pep3156,python-asyncio}.
Calling an \texttt{async def} function creates a coroutine.
Awaiting it executes the coroutine in the current task, while
\texttt{asyncio.create\_task} schedules it as a separate task
~\cite{pep492,python-asyncio-task}.
Tasks may be awaited, returned, or cancelled at different
program locations, making their lifetime depend on relationships
among operations on the same task.
Synchronous blocking operations delay other tasks on the event
loop~\cite{python-asyncio-dev}.
Task creation, \texttt{gather}, and queues can also admit work
that grows with input unless bounded.
Long-running operations commonly rely on timeouts or cancellation.
Cancellation is typically delivered as \texttt{CancelledError}
at an \texttt{await}; handlers and cleanup code determine
whether it propagates or execution continues~\cite{python-asyncio-task}.
These semantics make async risks depend on event order,
object relationships, and control flow through awaits,
cancellation, and scope exits.

\noindent
\textbf{CPython bytecode.}
CPython compiles Python source into code objects containing
instructions and execution metadata~\cite{python-dis}.
Bytecode instructions and exception metadata represent awaits,
suspension points, and exception-handling paths.
These representations vary across CPython versions.
For example, CPython~3.11 introduced exception tables
that changed how exception handling is encoded
~\cite{python-whatsnew-311}.

\subsection{Python Async in the Wild}\label{sec:bg:empirical}

\begin{figure}[htp]
  \centering
  \includegraphics[width=\linewidth]
    {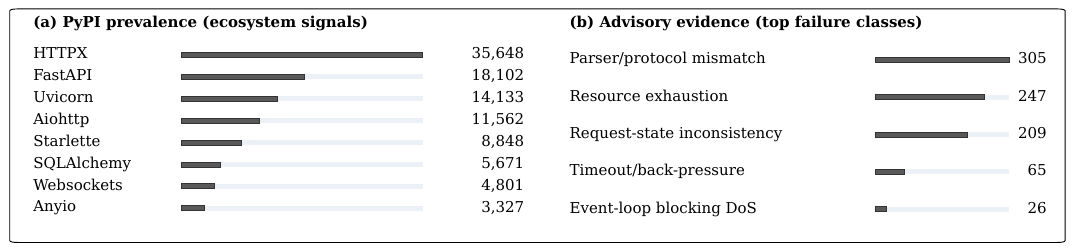}
  \vspace{-20pt}
  \caption{Python async ecosystem evidence.}
  \Description{Library-associated project counts and counts of the selected
  advisory failure categories in the motivating corpus.}
  \label{fig:motivation-data}
  \vspace{-10pt}
\end{figure}

We study PyPI packages released over a three-year period
~\cite{pypi-index-api,pypi-json-api}.
Source inspection identifies async code in 76{,}509 projects,
of which 76{,}453 provide source artifacts compatible with
CPython~3.10--3.14 and form our study population.
Figure~\ref{fig:motivation-data} shows project counts associated
with widely used Python async libraries.
We also collect public advisories concerning Python async
software~\cite{osv-dev,github-advisory-database,cve-program}.
After deduplication, we identify 1{,}787 distinct bugs.
Figure~\ref{fig:motivation-data} summarizes selected failure
categories, including resource exhaustion, request-state
inconsistency, and timeout-related failures.
These measurements show the prevalence of async code
and recurring failures in real Python software.

\subsection{Motivating Example}\label{sec:bg:motivating-example}

\begin{figure}[htp]
  \vspace{-0.5\baselineskip}
  \centering
  \includegraphics[width=\linewidth]
    {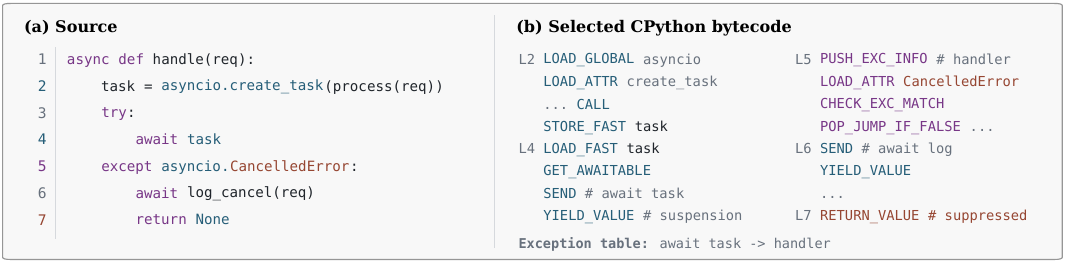}
  \vspace{-20pt}
  \caption{Task creation, awaiting, and cancellation suppression in
  source code and CPython~3.13 bytecode.}
  \Description{A Python async handler creates and awaits a task, catches
  cancellation at that await, performs asynchronous logging, and returns
  normally. The selected bytecode shows the task being stored and loaded,
  then the path from the suspension to the cancellation handler and
  normal return.}
  \label{fig:motivating-example}
  \vspace{-0.5\baselineskip}
\end{figure}

Figure~\ref{fig:motivating-example} shows an async request handler
that creates a task for \texttt{process(req)} and later awaits it.
If the handler is cancelled while \texttt{await task} is suspended,
cancellation enters the matching \texttt{CancelledError} handler.
The handler awaits \texttt{log\_cancel(req)} and returns normally
when logging completes, suppressing the cancellation.
The caller therefore observes normal completion rather than
cancellation, which can interfere with deadline or shutdown handling.

In this example, \textsc{flake8-async}~\cite{flake8-async}
can flag the \texttt{CancelledError} handler that returns
without propagating cancellation.
Its warning identifies the handler pattern but does not trace
how cancellation at \texttt{await task} reaches the handler
and leads to the normal return.
General-purpose Python call-graph and dataflow analyses can
recover calls and object relationships, but these facts alone
do not reconstruct the cancellation path.
Dynamic testing can reproduce the behavior when cancellation
reaches the suspension, but doing so depends on execution timing.
Analyzing this behavior statically presents four challenges.
CPython versions encode awaits and exception handling differently.
Recovering execution flow from the suspension to the handler
while preserving source locations requires version-aware
bytecode analysis (C1).
The task created at line~2 is awaited at line~4.
Relating these operations involves identifying their shared
symbolic task object (C2).
When cancellation occurs at the await, execution enters the
matching handler and reaches a normal return if logging completes.
Tracking this sequence requires preserving event order and
object state along the same execution path (C3).
Finally, the handler may return normally or propagate
\texttt{CancelledError}.
Distinguishing these outcomes and retaining their supporting
events and state facts enables risk detection and reporting (C4).

These challenges guide the design of {\tech}.
Version-aware bytecode processing reconstructs execution flow,
while the AETG connects async operations through symbolic objects.
Path-local simulation follows cancellation from the suspension
through handler execution, preserving event order and object state.
Risk-specific detectors then evaluate the resulting traces.
For the motivating example, the cancellation-suppression witness
connects the await, cancellation delivery, handler entry,
and normal return.

\section{Approach Overview}\label{sec:method}

\begin{figure*}[htp]
  \centering
  \includegraphics[width=0.96\textwidth]
    {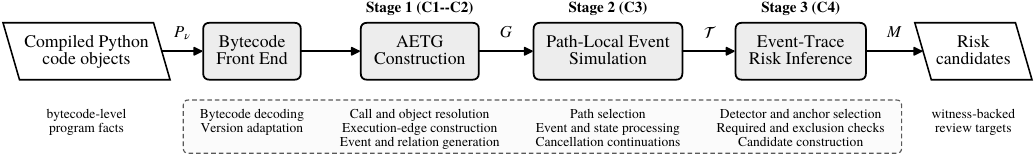}
  \vspace{-5pt}
  \caption{Overview of the {\tech} analysis workflow.}
  \Description{Compiled package code is transformed into an AETG, then
  simulated trace/state pairs, then risk candidates produced by detector plugins.}
  \label{fig:technical-workflow}
  \vspace{-10pt}
\end{figure*}

Given the code objects $P_\nu$ compiled from package $P$
under CPython version $\nu$, the bytecode front end decodes
instructions and exception-handling metadata while preserving
bytecode offsets and available source locations.
Figure~\ref{fig:technical-workflow} shows the three-stage
analysis workflow:
\(
P_{\nu} \xrightarrow{\mathcal{A}} G
\xrightarrow{\mathcal{S}} \mathcal{T}
\xrightarrow{\mathcal{I}_D} M .
\)
Here, $G$ is the \emph{Async Event Transition Graph} (AETG),
$\mathcal{T}$ contains execution traces and their abstract states,
and $M$ contains the reported risk candidates.
\emph{Stage 1: AETG Construction} resolves call targets and
symbolic objects, reconstructs bytecode execution flow,
and generates async events and relation edges.
These elements form the AETG $G$.
\emph{Stage 2: Path-Local Event Simulation} explores selected
execution paths in $G$, tracking event order and object states.
It also explores cancellation continuations from await suspensions,
producing trace/state pairs in $\mathcal{T}$.
\emph{Stage 3: Event-Trace Risk Inference} applies risk-specific
detectors to the resulting traces and states.
Each detector selects relevant anchors, evaluates its risk
conditions, and constructs candidates with supporting witnesses.
The resulting candidates form $M$.

\subsection{AETG Construction}\label{sec:aetg}

\subsubsection{AETG Model}

The \emph{Async Event Transition Graph} (AETG) represents
bytecode execution, async events, and relationships among operations.
For example, it connects task creation to the scheduled coroutine
and later awaits of the same task.

\begin{figure}[htp]
\centering
\begin{tcolorbox}[asyncbox,width=0.6\linewidth]
\scriptsize
\begin{minipage}[c]{0.32\linewidth}
\centering
$\displaystyle G=\langle V,E,X,\alpha\rangle$
\end{minipage}
\hfill
\begin{minipage}[c]{0.66\linewidth}
\[
\left\{
\begin{aligned}
v &= \langle co,off,op,loc\rangle\in V,\\[2pt]
e &= \langle v_i,v_j,k\rangle\in E,
     \quad k\in K_{\mathrm{exec}}\cup K_{\mathrm{rel}},\\[2pt]
x &= \langle kind,\bar{o}\rangle\in X(v),\\[2pt]
\alpha &:\ \text{derivation information}.
\end{aligned}
\right.
\]
\end{minipage}
\end{tcolorbox}
\vspace{-9pt}
\caption{AETG representation.}
\Description{The AETG consists of instruction nodes, typed execution
and relation edges, events attached to instructions, and their
derivation information.}
\label{fig:aetg-definition}
\vspace{-9pt}
\end{figure}

Figure~\ref{fig:aetg-definition} defines the AETG as
$G=\langle V,E,X,\alpha\rangle$.
Each node $v=\langle co,off,op,loc\rangle$ represents a
bytecode instruction, where $co$ identifies the code object,
$off$ the bytecode offset, $op$ the instruction kind,
and $loc$ the source location.
Each edge $\langle v_i,v_j,k\rangle\in E$ connects two
instruction nodes.
Edges with $k\in K_{\mathrm{exec}}$ represent possible
execution transfers, while edges with $k\in K_{\mathrm{rel}}$
connect related async operations or objects.
The corresponding edge sets are denoted by
$E_{\mathrm{exec}}$ and $E_{\mathrm{rel}}$.
$X(v)$ contains the events attached to node $v$.
Each event records an operation or control-flow point
and its associated operands.
For object operands such as tasks, points-to analysis provides
symbolic objects representing possible runtime objects.
A reference may point to several symbolic objects, and
operations sharing one may act on the same runtime object.
Finally, $\alpha$ records the source locations, resolved call
targets, object bindings, and operation rules used to derive
graph elements.

\subsubsection{Event and Relation Rules}

\leavevmode\par
\vspace{1pt}
\noindent
\blackcircleone{1} \emph{Event Generation.}
AETG events are generated from bytecode instructions and
resolved function calls.
Instruction semantics identify awaits and scope exits,
while $\Pi$ provides possible call targets, symbolic objects
used or returned, and receiver and argument bindings.
The following rules generate $\mathsf{TaskCreate}$ and
$\mathsf{Await}$ events:
\[
\frac{
  \substack{
    c\in\operatorname{Targets}_{\Pi}(v_c)
    \quad \operatorname{TaskFactory}(c)\\
    t\in\operatorname{Result}_{\Pi}(v_c)
  }
}{
  \langle\mathsf{TaskCreate},t\rangle\in X(v_c)
}
\qquad
\frac{
  \substack{
    \operatorname{AwaitSite}(v_a)\\
    o\in\operatorname{Operand}_{\Pi}(v_a)
  }
}{
  \langle\mathsf{Await},o\rangle\in X(v_a)
}.
\]
Here, $\operatorname{Targets}_{\Pi}(v_c)$ gives possible callees,
and $\operatorname{Result}_{\Pi}(v_c)$ gives the symbolic objects
returned by the call.
$\operatorname{TaskFactory}(c)$ holds when $c$ is a recognized
task-creation function, such as \texttt{asyncio.create\_task}.
$\operatorname{AwaitSite}(v_a)$ holds when $v_a$ corresponds to
an \texttt{await} operation, and
$\operatorname{Operand}_{\Pi}(v_a)$ gives the symbolic objects
being awaited.
An await may complete immediately or suspend execution.
Other recognized calls generate events for blocking operations,
task admission, queue admission, and external async operations.
For example, a resolved call to \texttt{time.sleep} generates
a blocking event.
Resolved queue operations identify the destination queue,
while constructor arguments provide known capacity bounds.
Operations inside analyzed callees generate events at the
corresponding instruction nodes.
For external calls without analyzed bytecode, matching operation
rules supply recognized behavior, while other effects are
conservatively approximated.

\vspace{1pt}
\noindent
\blackcircleone{2} \emph{Relation Edges.}
Relation edges connect instruction nodes through shared symbolic
objects, coroutine or callback entries, and execution reachability.
The following rules link task creation to awaits of the same
symbolic task and to the scheduled coroutine's entry:
\[
\frac{
  \substack{
    \langle\mathsf{TaskCreate},t\rangle\in X(v_c)\\
    \langle\mathsf{Await},t\rangle\in X(v_a)
  }
}{
  \langle v_c,v_a,\mathsf{task\_observed}\rangle\in E_{\mathrm{rel}}
}
\quad
\frac{
  \substack{
    \langle\mathsf{TaskCreate},t\rangle\in X(v_c)\\
    v_e\in\operatorname{Entries}_{\Pi}(t)
  }
}{
  \langle v_c,v_e,\mathsf{spawn}\rangle\in E_{\mathrm{rel}}
}.
\]
The $\mathsf{task\_observed}$ relation connects operations
involving the same symbolic task.
Recognized result-retrieval calls use the same object matching.
Simulation determines whether the task is observed along
each selected path.
For spawning, $\Pi$ associates the returned task with the
coroutine passed to the creation call.
$\operatorname{Entries}_{\Pi}(t)$ gives the possible entry nodes
of that coroutine, from which simulation explores the spawned task.
Other relations connect callback registrations to resolved
callback entries and task creation to returns carrying the task.
Execution reachability connects task creation to reachable
scope exits.
These edges identify possible relationships between operations,
which simulation evaluates along selected execution paths.

\subsubsection{Construction Procedure}

Given compiled code objects $P_\nu$,
{\tech} constructs the AETG in three steps:
call and object resolution, execution-edge construction,
and event and relation generation.
Algorithm~\ref{alg:aetg-construction} presents the procedure.
First, the bytecode front end decodes instructions and
exception metadata into $B$.
Call and object resolution uses points-to
analysis~\cite{andersen1994program} to identify possible
symbolic objects and call targets.
It also recovers receiver, argument, and result bindings,
collecting these facts in $\Pi$
(line~\ref{alg:aetg:resolve}).
Second, the analysis constructs instruction nodes and
execution edges from bytecode, including branches,
exception handling, cleanup, and await continuations
(lines~\ref{alg:aetg:nodes}--\ref{alg:aetg:exec}).
For available callee bodies, interprocedural edges connect
resolved synchronous calls and directly awaited coroutines
to their entry nodes.
Return edges connect normal completion to the caller or
awaiting computation.
Third, \texttt{bytecodeEvents} identifies operations from
instruction semantics, while \texttt{resolvedCallEvents}
uses resolved call targets and operand bindings in $\Pi$.
The resulting events are attached to instruction nodes
(line~\ref{alg:aetg:events}).
The rules defined above then generate relation edges
(line~\ref{alg:aetg:relations}).
Finally, $\alpha$ records the source locations, resolution
facts, and operation rules used to derive graph elements
(line~\ref{alg:aetg:provenance}).

\begingroup
\SetAlCapFnt{\footnotesize}
\SetAlCapNameFnt{\footnotesize\bfseries}
\begin{figure}[!t]
\centering
\begin{minipage}{0.7\linewidth}

\begin{algorithm}[H]
\footnotesize
\setlength{\algomargin}{1.4em}
\DontPrintSemicolon
\LinesNumbered
\SetAlgoNlRelativeSize{-1}
\caption{Construct the Async Event Transition Graph}
\label{alg:aetg-construction}

\SetKwFunction{Decode}{decodeBytecode}
\SetKwFunction{Resolve}{resolveCallAndObjectFacts}
\SetKwFunction{Nodes}{instructionNodes}
\SetKwFunction{Transfers}{recoverExecutionEdges}
\SetKwFunction{BytecodeEvents}{bytecodeEvents}
\SetKwFunction{CallEvents}{resolvedCallEvents}
\SetKwFunction{Relations}{deriveRelationEdges}
\SetKwFunction{Annotate}{attachProvenance}
\SetKw{Return}{return}

\KwIn{$P_\nu$: compiled package code objects}
\KwOut{$G=\langle V,E,X,\alpha\rangle$: AETG}

$B\gets\Decode(P_\nu)$\;

$\Pi\gets\Resolve(B)$
\nllabel{alg:aetg:resolve}\;

$V\gets\Nodes(B)$
\nllabel{alg:aetg:nodes}\;

$E_{\mathrm{exec}}\gets\Transfers(B,V,\Pi)$
\nllabel{alg:aetg:exec}\;

$X\gets\emptyset$\;
\ForEach{$v\in V$}{
    $X(v)\gets\BytecodeEvents(v,B,\Pi)\cup
    \CallEvents(v,\Pi)$
    \nllabel{alg:aetg:events}\;
}

$E_{\mathrm{rel}}\gets
\Relations(V,E_{\mathrm{exec}},X,\Pi)$
\nllabel{alg:aetg:relations}\;

$E\gets E_{\mathrm{exec}}\cup E_{\mathrm{rel}}$\;

$\alpha\gets\Annotate(B,V,E,X,\Pi)$
\nllabel{alg:aetg:provenance}\;

\Return{$\langle V,E,X,\alpha\rangle$}\;

\end{algorithm}
\end{minipage}
\Description{Pseudocode resolves calls and objects, constructs
instruction nodes and execution edges, generates async events
and relation edges, and records derivation information.}
\end{figure}
\endgroup

A bytecode CFG captures possible execution transfers, but
control flow alone does not show whether task creation and
a later await involve the same symbolic task.
The AETG therefore attaches async events to instruction nodes
and connects related operations through symbolic objects.
These relations identify possible task uses, while simulation
determines which operations occur along each execution path.
The derivation information in $\alpha$ links these graph
elements to their source locations and resolution facts.
Task creation also differs from a synchronous function call.
A synchronous call transfers control to its callee, whereas
task creation schedules a coroutine for separate execution.
Representing scheduling as a call edge would place the spawned
coroutine on the creator's execution path.
The $\mathsf{spawn}$ relation instead connects task creation
to the coroutine entry without merging their execution paths.
Simulation can therefore explore spawned tasks separately
while retaining their scheduling relationships.

\begin{figure*}[htp]
  \centering
  \includegraphics[width=0.97\textwidth]
    {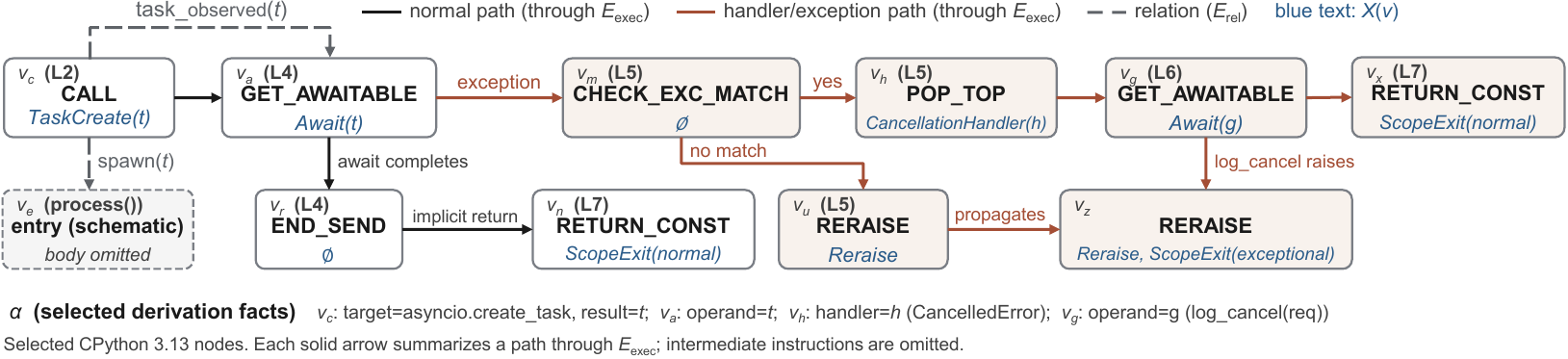}
  \vspace{-5pt}
  \caption{Selected AETG fragment for the motivating handler under
  CPython~3.13. Solid arrows represent execution paths, while
  dashed arrows represent task relations.}
  \Description{Instruction nodes contain selected async events.
  Solid arrows show normal and exceptional execution paths,
  while dashed arrows show task-observation and spawning relations.
  The alpha annotations record selected derivation facts.}
  \label{fig:aetg-motivation}
  \vspace{-10pt}
\end{figure*}

\noindent
\textbf{Running example.}
Figure~\ref{fig:aetg-motivation} shows the AETG fragment for
the handler in Figure~\ref{fig:motivating-example}.
Node $v_c$ records $\mathsf{TaskCreate}(t)$, and node $v_a$
records $\mathsf{Await}(t)$.
The shared symbolic task $t$ connects these nodes through a
$\mathsf{task\_observed}$ relation.
A $\mathsf{spawn}$ relation links $v_c$ to the entry of
\texttt{process}.
When an exception reaches node $v_m$, the exception check
directs \texttt{CancelledError} to handler $v_h$, while
other exceptions follow the re-raise path.
The graph also represents normal completion and exception
propagation from \texttt{log\_cancel}.
The annotations in $\alpha$ retain the resolved call targets,
object bindings, and exception-handling information used
to construct these elements.

\subsection{Path-Local Event Simulation}
\label{sec:phase2:symbolic-event-simulation}

Path-local event simulation explores selected execution paths
through the AETG, including cancellation continuations from
await suspensions.
It produces the event traces and abstract states used by
risk detectors.

\subsubsection{Traces and States}

For each completed path, the simulator records an event trace
and its final abstract state.
The trace contains events in execution order, including
cancellation events introduced during simulation:
\[
(\pi,\sigma_\pi)\in\mathcal{T},
\quad
\pi=\langle x_1,\ldots,x_n\rangle,
\quad
x_i=\langle kind,site,obj,prov\rangle.
\]
Each occurrence $x_i$ records the operation or control-flow
point in $kind$, its AETG node in $site$, and any associated
symbolic object in $obj$.
The field $prov$ records the source and resolution facts
supporting the event.
For occurrences $x_i$ and $x_j$ in $\pi$,
$x_i\prec_\pi x_j$ means that $i<j$, and
$x_i\equiv_o x_j$ means that both refer to the same symbolic object.
As the simulator traverses path $p$, it appends events to the
trace and updates the abstract state using the transfer
function $\delta$:
\[
\sigma_p=\langle T,F,H,\Phi\rangle,
\qquad
\delta(\sigma_p,x_i)=\sigma'_p.
\]
The map $T$ records properties of symbolic tasks.
For task $t$,
\[
T(t)\subseteq
\{\mathsf{pending},\mathsf{owned},\mathsf{observed},\mathsf{escaped}\}.
\]
A task becomes $\mathsf{observed}$ when an operation retrieves
its result or exception, while a completed task may remain unobserved.
$F$ records task and queue admissions and their known bounds,
while $H$ records synchronous calls.
$\Phi$ records execution facts used by the detectors, such as
cancellation reaching a handler or being reraised.
After processing the final event on path $p$, the simulator
stores $\sigma_p$ as $\sigma_\pi$ and adds
$(\pi,\sigma_\pi)$ to $\mathcal{T}$.
The summary $\Sigma$ collects event types and state properties
across the completed traces.
Risk inference uses $\Sigma$ to select detectors, which then
evaluate individual trace/state pairs.

\subsubsection{Simulation Procedure}
\label{sec:phase2:procedure}

Given an AETG and path budget $N_p$, the simulator selects
paths, processes events, and explores cancellation continuations.
Algorithm~\ref{alg:event-trace-generation} presents the procedure.

\vspace{1pt}
\noindent
\blackcircleone{1} \emph{Path selection.}
The simulator starts from the entry nodes of analyzed package
code objects.
Spawn and callback-registration relations provide additional
roots for separate coroutine and callback traversals.
From these roots, it selects at most $N_p$ completed paths
through $E_{\mathrm{exec}}$ using deterministic depth-first search
(line~\ref{alg:sim:paths}).
At each branch, it prioritizes successors leading to previously
unexplored nodes or execution edges.
Successors with equal priority follow a fixed node order.
A path traverses a given static execution edge at most once.
This limits repeated loop iterations and visits to shared
callee bodies.
Path prefixes stopped by this restriction are discarded.
Only paths reaching a normal or exceptional scope exit
are retained for simulation.

\vspace{1pt}
\noindent
\blackcircleone{2} \emph{Event and state processing.}
For each selected path, the simulator processes instruction events
and activates applicable relation edges
(line~\ref{alg:sim:events}).
It appends the resulting events to the trace in traversal order
and applies $\delta$ to update the abstract state.
For example, task creation adds $\mathsf{pending}$ to $T(t)$.
A $\mathsf{task\_observed}$ relation connects task creation to
awaits or result-retrieval calls involving the same symbolic task.
When an operation retrieves the task's result or exception,
the simulator updates its state with:
\[
T'(t)=T(t)\cup\{\mathsf{observed}\}.
\]
At the end of the path, the simulator records the normal or
exceptional scope exit
(line~\ref{alg:sim:termination}).
Detectors use the resulting trace and state to evaluate
conditions at that exit.

\vspace{1pt}
\noindent
\blackcircleone{3} \emph{Cancellation.}
When a path reaches a suspension at an \texttt{await},
the simulator explores an additional execution in which
cancellation is delivered at that point.
It copies the trace prefix and abstract state at the suspension,
preserving the operations and task states recorded before
cancellation
(line~\ref{alg:sim:cancel}).
The simulator appends $\mathsf{CancelAt}(a)$ to the copied trace,
where $a$ identifies the await node, and associates the delivery
with the specific await occurrence.
Applying $\delta$ records delivery of \texttt{CancelledError}.
The simulator follows exception and cleanup edges in the AETG,
selects handlers according to the active exception, and explores
each possible branch with a separate trace and state.
At handler entry, it records which delivered cancellation reaches
the handler.
It then processes subsequent events using the same transfer
function, recording whether cancellation propagates or
execution continues normally.
The recorded await and handler associations provide
$\mathsf{AtAwait}$ and $\mathsf{Handles}$ for Stage~3.
Each continuation contains one injected cancellation.
The path prefix and continuation together traverse a given
static execution edge at most once.
Later awaits follow the existing execution and exception edges.
Only continuations reaching a normal or exceptional scope exit
contribute trace/state pairs.
The original path and its completed cancellation continuations
produce separate trace/state pairs.
Since $N_p$ limits original path selection, $\mathcal{T}$ may
contain more than $N_p$ pairs.
Finally, $\Sigma$ summarizes the resulting event types and
state properties
(line~\ref{alg:sim:summary}).

\begingroup
\SetAlCapFnt{\footnotesize}
\SetAlCapNameFnt{\footnotesize\bfseries}
\begin{figure}[!t]
\centering
\begin{minipage}{0.7\linewidth}

\begin{algorithm}[H]
\footnotesize
\setlength{\algomargin}{1.4em}
\DontPrintSemicolon
\LinesNumbered
\SetAlgoNlRelativeSize{-1}
\caption{Generate Path-Local Event Traces}
\label{alg:event-trace-generation}

\SetKwFunction{GetRoots}{analysisRoots}
\SetKwFunction{Select}{selectPaths}
\SetKwFunction{Process}{processEvents}
\SetKwFunction{Cancel}{exploreCancellation}
\SetKwFunction{Finish}{recordTermination}
\SetKwFunction{Summarize}{summarize}
\SetKw{Return}{return}

\KwIn{$G$: AETG; $N_p$: path budget}
\KwOut{$\mathcal{T}$: trace/state pairs;
       $\Sigma$: trace summary}

$\mathcal{T}\gets\emptyset$\;

$Roots\gets\GetRoots(G)$\;

$\mathcal{P}\gets\Select(G,Roots,N_p)$
\nllabel{alg:sim:paths}\;

\ForEach{$p\in\mathcal{P}$}{
    $\pi\gets\langle\rangle$;
    $\rho\gets\langle\rangle$;
    $\sigma\gets\operatorname{initState}()$\;

    \ForEach{node or edge $q$ in $p$}{
        $\rho\gets\rho\cdot q$\;

        $(\pi,\sigma)\gets
        \Process(G,q,\rho,\pi,\sigma)$
        \nllabel{alg:sim:events}\;

        \If{$q$ is a suspension at await $a$}{
            $\mathcal{C}_a\gets
            \Cancel(G,a,\rho,\pi,\sigma)$
            \nllabel{alg:sim:cancel}\;

            $\mathcal{T}\gets\mathcal{T}\cup\mathcal{C}_a$\;
        }
    }

    $\sigma\gets\Finish(G,p,\sigma)$
    \nllabel{alg:sim:termination}\;

    $\mathcal{T}\gets\mathcal{T}\cup\{(\pi,\sigma)\}$\;
}

$\Sigma\gets\Summarize(\mathcal{T})$
\nllabel{alg:sim:summary}\;

\Return{$(\mathcal{T},\Sigma)$}\;

\end{algorithm}
\end{minipage}
\Description{Pseudocode selects completed execution paths,
processes instruction and relation events, explores completed
cancellation continuations, and summarizes the resulting
trace/state pairs.}
\end{figure}
\endgroup

The AETG may connect task creation to an observation on one branch
and a scope exit on another.
Combining states across these branches could make the task appear
observed on a path that exits without observing it.
The simulator therefore maintains a separate trace and abstract
state for each explored path.
The trace preserves event order, while the state records how
operations affect symbolic tasks on that path.
This allows detectors to check task observation and ownership
at the scope exit reached by each path.
Cancellation may reach the same handler from different await
suspensions.
Starting simulation at the handler would lose the execution
history identifying where cancellation was delivered.
The simulator therefore copies the trace and state at the
suspension and follows the exception and cleanup paths.
This connects cancellation delivery to the corresponding handler
and records whether cancellation propagates or execution
continues normally.
Keeping the original path and cancellation continuation separate
preserves their different outcomes.

\vspace{1pt}
\noindent
\textbf{Running example.}
Figure~\ref{fig:aetg-motivation} shows a path that creates
task $t$, awaits it, and returns normally.
At the suspension in \texttt{await task}, the simulator also
explores a cancellation continuation.
This continuation enters the matching handler, awaits
\texttt{log\_cancel(req)}, and returns normally if logging completes.
The following traces show selected events and their node
and object fields.
Here, $c$ is the task-creation node,
$a_{\mathit{task}}$ and $a_{\mathit{log}}$ are the await nodes,
$h$ identifies the handler occurrence, and $g$ is the symbolic
awaitable returned by \texttt{log\_cancel(req)}.
$\mathsf{CancellationHandlerContinues}(h)$ records the handler's
normal continuation:
\[
\begin{aligned}
\pi_{\mathit{normal}}=\langle
&\ldots,
 \mathsf{TaskCreate}(c,t),
 \mathsf{Await}(a_{\mathit{task}},t),
 \ldots,
 \mathsf{ScopeExit}(\mathsf{normal})\rangle,\\[2pt]
\pi_{\mathit{cancel}}=\langle
&\ldots,
 \mathsf{TaskCreate}(c,t),
 \mathsf{Await}(a_{\mathit{task}},t),
 \mathsf{CancelAt}(a_{\mathit{task}}),\\
&\mathsf{CancellationHandler}(h),
 \mathsf{Await}(a_{\mathit{log}},g),
 \ldots,\\
&\mathsf{CancellationHandlerContinues}(h),
 \mathsf{ScopeExit}(\mathsf{normal})\rangle.
\end{aligned}
\]
Both traces reach a normal scope exit.
The cancellation trace additionally records cancellation delivery
and the handler's normal continuation, providing the event
sequence used to detect cancellation suppression.

\subsection{Event-Trace Risk Inference}
\label{sec:phase3:event-trace-risk-inference}

Event-trace risk inference evaluates completed execution traces
and their abstract states using risk-specific detector plugins.
Each plugin reports candidates with supporting witnesses.

\subsubsection{Detector Interface}

Each detector plugin is defined as
$d_i=\langle id_i,Q_i,P_i,W_i\rangle$.
The identifier $id_i$ names the detector,
$Q_i$ selects event or object anchors,
$P_i$ checks the risk conditions, and
$W_i$ constructs a candidate and its witness.
For an anchor $\eta$ and trace/state pair $(\pi,\sigma_\pi)$,
a candidate is generated when $\mathsf{Match}_i$ holds:
\[
\begin{aligned}
\mathsf{Match}_i(\eta,\pi,\sigma_\pi)
&\equiv
\eta\in Q_i(\pi,\sigma_\pi)
\land P_i(\eta,\pi,\sigma_\pi),\\[2pt]
m
&=W_i(\eta,\pi,\sigma_\pi)
 =\langle id_i,loc,\eta,witness,prov\rangle.
\end{aligned}
\]
The candidate records the detector identifier, source location
$loc$, anchor $\eta$, witness, and derivation facts in $prov$.
The witness contains the event occurrences and state facts
that establish the match.

\begingroup
\SetAlCapFnt{\footnotesize}
\SetAlCapNameFnt{\footnotesize\bfseries}
\begin{figure}[htp]
\centering
\begin{minipage}{0.7\linewidth}

\begin{algorithm}[H]
\footnotesize
\setlength{\algomargin}{1.4em}
\DontPrintSemicolon
\LinesNumbered
\SetAlgoNlRelativeSize{-1}
\caption{Infer Risks from Event Traces}
\label{alg:event-trace-risk-inference}

\SetKwFunction{MayApply}{mayApply}
\SetKwFunction{Select}{selectAnchors}
\SetKwFunction{Check}{checkConditions}
\SetKwFunction{Build}{buildCandidate}
\SetKwFunction{Canonicalize}{canonicalize}
\SetKw{Continue}{continue}
\SetKw{Return}{return}

\KwIn{$\mathcal{T}$: trace/state pairs;
       $\Sigma$: trace summary;
       $D$: detector plugins}
\KwOut{$M$: risk candidates}

$M\gets\emptyset$\;

\ForEach{$d_i\in D$}{
    \If{$\Sigma\neq\emptyset
        \land\neg\MayApply(d_i,\Sigma)$}{
        \Continue
        \nllabel{alg:infer:prefilter}\;
    }

    \ForEach{$(\pi,\sigma_\pi)\in\mathcal{T}$}{
        $A\gets\Select(d_i,\pi,\sigma_\pi)$
        \nllabel{alg:infer:anchors}\;

        \ForEach{$\eta\in A$}{
            \If{$\Check(d_i,\eta,\pi,\sigma_\pi)$
                \nllabel{alg:infer:predicate}}{
                $m\gets\Build(d_i,\eta,\pi,\sigma_\pi)$
                \nllabel{alg:infer:witness}\;

                $M\gets M\cup\{\Canonicalize(m)\}$
                \nllabel{alg:infer:canonicalize}\;
            }
        }
    }
}

\Return{$M$}\;

\end{algorithm}
\end{minipage}
\Description{Pseudocode selects detectors and anchors, checks risk
conditions, constructs candidates with witnesses, and removes
duplicate reports.}
\end{figure}
\endgroup

\subsubsection{Inference Procedure}

Given the completed trace/state pairs $\mathcal{T}$,
their summary $\Sigma$, and detector plugins $D$,
the inference procedure generates risk candidates.
Algorithm~\ref{alg:event-trace-risk-inference} presents the procedure.
The analysis first uses $\Sigma$ to select detectors whose
required event or state information may be present
(line~\ref{alg:infer:prefilter}).
An empty summary leaves all detectors eligible.
For each trace/state pair, a selected detector identifies
relevant event occurrences or symbolic objects as anchors
(line~\ref{alg:infer:anchors}).
For each anchor, \texttt{checkConditions} evaluates the
detector's risk predicate
(line~\ref{alg:infer:predicate}).
When the predicate holds, \texttt{buildCandidate} constructs
a candidate with its supporting witness
(line~\ref{alg:infer:witness}).
The final step canonicalizes candidates to remove duplicate
reports
(line~\ref{alg:infer:canonicalize}).

A trace may contain several tasks and exception handlers,
but each risk concerns specific operations and their relationships.
For task lifetime, a creation event alone does not establish a risk;
the detector examines the task's observation and ownership state
when its creator exits.
For cancellation suppression, a handler's normal continuation
matters when cancellation delivered at an earlier await reaches
that handler.
The detector therefore selects the relevant task or handler
as an anchor and evaluates its risk conditions using the
corresponding trace and state.
This also keeps different outcomes on separate execution paths
from being combined.
Different risk families use different combinations of events,
object relationships, and state facts.
The shared interface allows detectors to evaluate these conditions
without repeating graph construction or path simulation.
For each match, the detector retains the relevant events,
state facts, and source locations in a witness.
The witness connects the reported condition to the execution
history used during source review.

\vspace{1pt}
\noindent
\textbf{Example plugin:}
\emph{Cancellation suppression} (\texttt{AEGISASYNC003}).
This detector examines handlers reached by cancellation
delivered at an await.
The anchor query uses the associations recorded during simulation.
$\mathsf{AtAwait}(x_j,x_i)$ holds when cancellation event $x_j$
was introduced at await occurrence $x_i$.
$\mathsf{Handles}(x_k,x_j)$ holds when handler-entry occurrence
$x_k$ receives that cancellation.
The query and risk predicate are defined as follows:
\[
\begin{aligned}
Q_{\mathit{suppress}}(\pi,\sigma_\pi)
=\{\,x_k\in\pi\mid {}&
\exists x_i,x_j\in\pi.\;
x_i\prec_\pi x_j\prec_\pi x_k\\
&{}\land\mathsf{AtAwait}(x_j,x_i)
\land\mathsf{Handles}(x_k,x_j)\,\},\\[2pt]
P_{\mathit{suppress}}(x_k,\pi,\sigma_\pi)
\equiv {}&
\bigl(
\sigma_\pi\models
\mathsf{CancellationHandlerContinues}(x_k)
\bigr)\\
&{}\land\neg\bigl(
\exists x_\ell\in\pi.\;
x_k\prec_\pi x_\ell
\land\mathsf{CancellationReraise}(x_\ell,x_k)
\bigr)\\
&{}\land\neg\bigl(
\sigma_\pi\models
\mathsf{CancellationHandlerLoopExit}(x_k)
\bigr).
\end{aligned}
\]
Here, $\sigma_\pi\models f$ means that the state records
fact $f$.
$\mathsf{CancellationHandlerContinues}(x_k)$ records that
the handler returns normally or execution continues after it.
$\mathsf{CancellationReraise}(x_\ell,x_k)$ identifies an event
that propagates the cancellation handled by $x_k$.
$\mathsf{CancellationHandlerLoopExit}(x_k)$ records a loop-exit
condition recognized for that handler.
The witness function $W_{\mathit{suppress}}$ retains the await,
cancellation delivery, handler entry, and handler outcome,
together with their source locations and resolution facts.

\vspace{1pt}
\noindent
\textbf{Running example.}
For the motivating handler, the detector evaluates
$(\pi_{\mathit{cancel}},\sigma_{\pi_{\mathit{cancel}}})$
from Section~\ref{sec:phase2:symbolic-event-simulation}.
$Q_{\mathit{suppress}}$ selects the handler-entry occurrence
$x_k$ reached by cancellation.
The recorded normal return satisfies
$\mathsf{CancellationHandlerContinues}(x_k)$.
The trace contains no re-raise of the handled cancellation,
and the state records no recognized loop exit.
Thus, $P_{\mathit{suppress}}$ holds, and
$W_{\mathit{suppress}}$ constructs an
\texttt{AEGISASYNC003} candidate.
Its witness connects the await, cancellation delivery,
handler entry, and normal return in the same execution trace.

\section{Implementation and Limitations}
\label{sec:implementation}

{\tech} is implemented in Python with 15.8~KLoC and supports
CPython~3.10--3.14.

\subsection{Implementation of Key Components}
\label{sec:implementation:components}

\vspace{1pt}
\noindent
\blackcircleone{1} \emph{Bytecode front end and version adapters.}
The front end invokes the target CPython interpreter to compile
package source and traverses the resulting nested code objects.
It uses \texttt{dis} to decode instructions and exception-handling
metadata~\cite{python-dis}.
Version adapters map opcode and exception-handling formats to
common instruction categories.
Instruction records retain operands, jump targets, stack effects,
bytecode offsets, and available source locations.

\vspace{1pt}
\noindent
\blackcircleone{2} \emph{AETG construction.}
The AETG builder constructs a package-level graph from
instruction-level CFGs and resolved interprocedural edges.
An inclusion-based worklist solver implements Andersen-style
points-to analysis~\cite{andersen1994program}.
It propagates possible call targets and symbolic objects through
imports, aliases, arguments, returns, and supported attribute accesses.
Relative imports are resolved using package and module metadata.
Recognition rules identify supported async operations from resolved
calls and their operand bindings.

\vspace{1pt}
\noindent
\blackcircleone{3} \emph{Path-local event simulation.}
The simulator uses deterministic depth-first search with a default
budget of $N_p=4096$ original paths per package-version artifact.
It prioritizes successors reaching previously unexplored nodes
or execution edges and uses a fixed node order to resolve ties.
A path traverses a given static execution edge at most once.
Prefixes stopped by this restriction are discarded.
At await suspensions, the simulator explores cancellation
continuations from copies of the current trace and abstract state.
The same edge restriction applies across the path prefix and
its continuation.
Only paths reaching normal or exceptional scope exits contribute
trace/state pairs.
The resulting traces and states are summarized in $\Sigma$
for detector selection.

\vspace{1pt}
\noindent
\blackcircleone{4} \emph{Event-trace risk inference.}
The detector registry contains six plugins:
\emph{Task lifetime}, \emph{Blocking operation},
\emph{Admission/fan-out}, \emph{Cancellation suppression},
\emph{Queue admission}, and \emph{Missing timeout}.
The registry uses the trace summary $\Sigma$ to select applicable
plugins before evaluating the trace/state pairs.
Each plugin implements anchor selection, risk-condition checking,
and witness construction through the shared interface.
The implementation stores candidates in a common report format
containing package and source metadata, the detector identifier,
supporting events and state facts, and derivation information.
Candidate records are canonicalized to remove duplicate reports.

\subsection{Limitations}
\label{sec:implementation:limitations}

{\tech} explores spawned tasks from separate roots and does not
model task interleavings or cross-task happens-before relationships.
A path traverses a given static execution edge at most once,
which can omit repeated loop iterations and calls through
the same callee edges.
Prefixes stopped by this restriction are discarded,
and the path budget further limits exploration.
These restrictions can leave some execution behaviors unanalyzed.
The points-to analysis is field-, context-, and flow-insensitive.
Object bindings can combine targets from different
fields, call contexts, and assignments.
Dynamic attributes and container aliases can also leave object
relationships unresolved.
The simulator tracks a finite set of task and execution properties.
Branch conditions involving untracked values may remain unresolved,
allowing exploration of runtime-infeasible paths.

For external calls without available bytecode, the analysis
conservatively approximates behavior using recognized operation
rules and available resolution facts.
Internal behavior outside these rules may be absent from the
generated events and traces.
Missing dependencies, native code, and generated code can
further limit the recovered behavior.
Support for context-manager cleanup and async generators remains
partial.
Structured-concurrency libraries and custom event loops may use
ownership or cancellation policies not represented by the analysis.
The six implemented detectors cover the evaluated risk families
but do not include dedicated cleanup-bypass detection.
Merged object information and infeasible paths can produce false
positives, while omitted code, paths, or operations can produce
false negatives.

\section{Evaluation}\label{sec:results}

We evaluate {\tech} through four research questions:

\noindent\textbf{RQ1.} Does {\tech} preserve async evidence and
accurately detect labeled risks?

\noindent\textbf{RQ2.} Can {\tech} analyze real-world packages
reliably and efficiently?

\noindent\textbf{RQ3.} How do path budget and detector count
affect effectiveness and cost?

\noindent\textbf{RQ4.} What candidates are reported in real-world
packages, and how many are confirmed?

\subsection{Evaluation Setup}\label{sec:eval:setup}

\noindent
\textbf{Experimental environment.}
We run all experiments on a 64-bit Ubuntu~22.04 workstation
with an Intel Core i9-13900F processor and 64~GB RAM.
We use separate environments for CPython~3.10--3.14
and analyze each package in a fresh process under its
assigned interpreter.

\noindent
\textbf{AsyncBench.}
We construct AsyncBench to evaluate detection under controlled
conditions.
The cases draw on failure patterns in advisories and documented
Python async semantics
~\cite{python-asyncio,python-asyncio-task,python-asyncio-dev}.
The benchmark and detector taxonomy draw on some of the same
advisory and semantic sources.
The benchmark manifest records risk families, source locations,
required witness facts, and expected outcomes.
Mappings to {\tech} plugins are stored separately from
the benchmark labels.
AsyncBench contains 78 cases and 180 labeled risk instances,
with 30 instances in each of the six risk families.
The cases include single and multiple risks and exercise aliasing,
interprocedural flow across functions and modules, cleanup,
queues, and timeouts.
Safe controls exercise the same features.
We evaluate every case under CPython~3.10--3.14.
For each labeled risk and tested version, we check whether
the analysis preserves the source locations, AETG relations,
events, and witness facts required by the manifest.

\noindent
\textbf{Package corpus and sampling.}
We use PyPI source artifacts from the same three-year period
as the motivation study.
Source inspection confirms async code in each artifact.
The population contains 76,453 artifacts, with one selected
release per PyPI project.
We use Python-version compatibility information to assign
each artifact to one compatible CPython version from 3.10 to 3.14.
The corpus manifest records the deterministic assignment rule.
We stratify the population by assigned interpreter version
and compiled instruction count.
Within each version group, we divide artifacts into ten
instruction-count deciles and sample 50 from each decile
using a fixed seed.
This yields 500 artifacts per version and 2,500 in total.
Table~\ref{tab:package-population-sample} reports the population,
sampling rates, and characteristics of the analyzed artifacts.
The package-scale results describe this stratified sample.

\begin{table*}[htp]
\centering
\caption{Package corpus, stratified sample, and analysis characteristics.}
\vspace{-10pt}
\label{tab:package-population-sample}
\smallskip
\small
\setlength{\tabcolsep}{5pt}

\resizebox{0.9\textwidth}{!}{%
\begin{tabular}{@{}lrrr rr rr@{}}
\toprule
& \multicolumn{3}{c}{Corpus and sampling}
& \multicolumn{4}{c}{Completed analyses} \\
\cmidrule(lr){2-4}\cmidrule(lr){5-8}
CPython
& Artifacts
& Sampled
& Sampling rate
& Python files
& Async code objects
& \multicolumn{2}{c}{Instructions per artifact} \\
\cmidrule(lr){7-8}
& & & & & & Median & P95 \\
\midrule
3.10 & 50,039 & 500 & 1.0\%  & 24,210 & 19,922 & 5,921  & 105,214 \\
3.11 & 15,090 & 500 & 3.3\%  & 25,583 & 30,886 & 9,522  & 105,823 \\
3.12 & 8,054  & 500 & 6.2\%  & 21,650 & 26,274 & 7,711  & 106,053 \\
3.13 & 2,512  & 500 & 19.9\% & 24,622 & 39,342 & 6,844  & 101,392 \\
3.14 & 758    & 500 & 66.0\% & 25,222 & 34,920 & 10,248 & 155,232 \\
\midrule
Total & 76,453 & 2,500 & 3.3\% & 121,287 & 151,344 & 7,842 & 112,391 \\
\bottomrule
\end{tabular}%
}
\end{table*}

\noindent
\textbf{Baseline and metrics.}
We run \textsc{flake8-async}~\cite{flake8-async} on the same
2,500 artifacts to compare analysis cost and detection
in risk families covered by both tools.
We measure analysis time per artifact and peak Python heap
usage within each profiled phase.
Simulation and inference are profiled together
(Stages~2 and~3).
The detector-count experiment measures inference cost
using precomputed traces.
%
On AsyncBench, a {\tech} report is counted as a true positive
when its risk family and source location match a labeled
risk instance and its witness contains the required event evidence.

\subsection{RQ1: Evidence Preservation and Effectiveness}
\label{sec:eval:effectiveness}

RQ1 evaluates whether {\tech} preserves the evidence required
by its detectors across CPython~3.10--3.14 and accurately
identifies the labeled risks in AsyncBench.

\begin{table*}[htp]
\centering
\begin{minipage}[t]{0.48\textwidth}
\vspace{0pt}
\centering
\caption{AsyncBench evidence preservation.}
\vspace{-10pt}
\label{tab:asyncbench-evidence}
\smallskip
\scriptsize
\setlength{\tabcolsep}{3pt}
\renewcommand{\arraystretch}{1.10}

\begin{tabularx}{\linewidth}{@{}
>{\raggedright\arraybackslash}Xlr@{}}
\toprule
Check & Unit & Result \\
\midrule
Analysis completion
  & Case-version runs & 390/390 \\
Classification consistency
  & Benchmark cases & 78/78 \\
Source-location preservation
  & Risk-version instances & 900/900 \\
AETG relation recovery
  & Risk-version instances & 900/900 \\
Async-event generation
  & Risk-version instances & 900/900 \\
Required witness facts
  & Risk-version instances & 900/900 \\
Paths per case-version
  & Median/P95/Max & 5/59/344 \\
\bottomrule
\end{tabularx}

\smallskip
\raggedright
Results cover CPython~3.10--3.14.
Risk-version counts record instances satisfying the required
manifest checks.
\end{minipage}
\hfill
\begin{minipage}[t]{0.49\textwidth}
\vspace{0pt}
\centering
\caption{AsyncBench detection results.}
\vspace{-10pt}
\label{tab:asyncbench-effectiveness}
\smallskip
\scriptsize
\setlength{\tabcolsep}{3pt}
\renewcommand{\arraystretch}{1.10}

\begin{tabularx}{\linewidth}{@{}
>{\raggedright\arraybackslash}Xrrr@{}}
\toprule
Family & Risks & {\tech} & \textsc{flake8-async} \\
\midrule
Task lifetime
  & 30 & 30/0/0 & 11/0/19 \\
Blocking operation
  & 30 & 30/0/0 & 5/0/25 \\
Admission/fan-out
  & 30 & 30/0/0 & -- \\
Cancellation suppression
  & 30 & 30/0/0 & 24/0/6 \\
Queue admission
  & 30 & 30/0/0 & -- \\
Missing timeout
  & 30 & 30/0/0 & -- \\
\midrule
Shared families
  & 90 & 90/0/0 & 40/0/50 \\
Full benchmark
  & 180 & 180/0/0 & -- \\
\bottomrule
\end{tabularx}

\end{minipage}
\end{table*}

\vspace{3pt}
\noindent
\blackcircleone{1} \emph{Evidence preservation.}
Table~\ref{tab:asyncbench-evidence} shows that all 390
case-version runs complete successfully, with consistent
classifications for all 78 cases across five CPython versions.
For all 900 risk-version instances, {\tech} preserves the
required source locations, AETG relations, async events,
and simulation witnesses.
The simulator explores a median of 5 paths per case-version,
with a P95 of 59 and a maximum of 344, all below the
default budget $N_p=4096$.
These results test whether the three analysis stages preserve
the information required by the detectors across different
bytecode versions.
The AETG recovers the relevant operations and object relations,
while simulation preserves their execution order and associated
state facts.
The complete evidence checks and consistent classifications
show that these detector inputs are preserved across all
five tested versions.

\vspace{3pt}
\noindent
\blackcircleone{2} \emph{End-to-end effectiveness.}
Table~\ref{tab:asyncbench-effectiveness} shows that {\tech}
detects all 180 labeled risks with zero false positives
on each CPython version.
The detections cover single-risk, mixed-risk, and
systematic-stress cases, while the 14 safe controls
produce no reports.
{\tech} achieves 100\% precision and recall on AsyncBench.
For the 90 risks in the three families covered by both tools,
{\tech} detects all 90, compared with 40 for
\textsc{flake8-async}.
The baseline achieves 100\% precision and 44.4\% recall.
Of its 50 missed instances, 44 occur in Task lifetime
and Blocking operation, with 19 and 25 misses, respectively.
The remaining 6 involve Cancellation suppression.
These differences exercise three capabilities of {\tech}.
For Task lifetime, symbolic-object relations connect task
creation to later uses, while path-local states record
observation and ownership at scope exit.
For Blocking operation, call resolution identifies blocking
effects across aliases and analyzed function bodies.
For Cancellation suppression, simulation connects cancellation
delivery to handler execution and its resulting outcome.
The corresponding detector predicates use these facts
to identify all 50 instances missed by the baseline.
Appendix~\ref{app:rq1-structural} reports results by case structure.

\find{Across CPython~3.10--3.14, {\tech} preserves the required
AETG and simulation evidence for all 180 AsyncBench risks
and detects them with 100\% precision and recall.
In the three shared families, it identifies all 50 instances
missed by \textsc{flake8-async}, exercising symbolic-object
relations, resolved call effects, and cancellation paths.}
\subsection{RQ2: Applicability and Efficiency}
\label{sec:eval:efficiency}

RQ2 evaluates analysis completion and cost on real-world packages,
compares {\tech} with \textsc{flake8-async}, and examines
how analysis cost changes with package size.
Component profiles identify the time and memory used by
the three analysis stages.

\begin{figure*}[htp]
\centering
\includegraphics[width=0.92\textwidth]
  {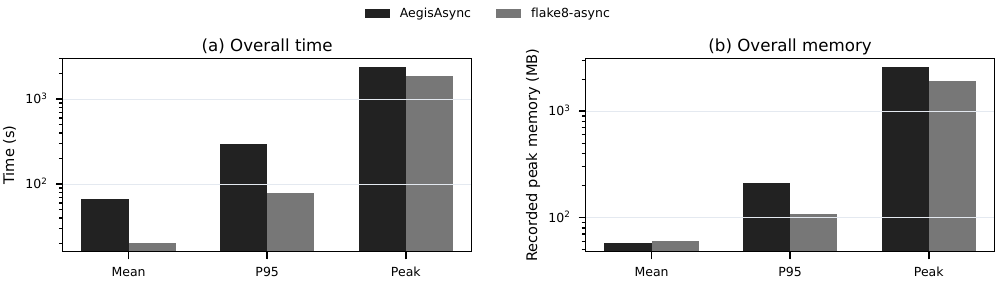}
\vspace{-10pt}
\caption{Package-level analysis cost on sampled artifacts:
(a) analysis time and (b) recorded peak memory.}
\label{fig:rq2-overall-comparison}
\Description{Analysis time and recorded peak-memory statistics
for AegisAsync and flake8-async.
AegisAsync has higher mean and P95 analysis time.
Recorded memory values have similar means, with higher
P95 and maximum values for AegisAsync.}
\end{figure*}

\vspace{2pt}
\noindent
\blackcircleone{1} \emph{Applicability and overall cost.}
{\tech} completes analysis on all 2,500 sampled artifacts,
covering 121,287 Python files, 73.1 million bytecode instructions,
and 151,344 async code objects.
The resulting AETGs contain 13.6 million edges.
At the default budget $N_p=4096$, the simulator explores
2,697,052 event paths, averaging 1,079 per artifact.
The path budget is reached in 188 analyses (7.5\%).
Thus, graph construction and path simulation complete across
the sample, with most analyses remaining below the configured
path budget.
%
Figure~\ref{fig:rq2-overall-comparison}(a) shows that {\tech}
takes 67.17~s per artifact on average, compared with 20.26~s
for \textsc{flake8-async}.
The corresponding P95 times are 293.97~s and 79.76~s.
{\tech} therefore takes 3.3$\times$ the baseline's mean
analysis time and 3.7$\times$ its P95 time.
Figure~\ref{fig:rq2-overall-comparison}(b) shows a different
memory pattern.
Both tools have similar mean recorded peak memory, but
{\tech} has higher P95 and maximum values.
Its mean, P95, and maximum peak Python heap are
57.5~MB, 211.0~MB, and 2,599.9~MB.
The larger upper-tail values show that some artifacts
require substantially more memory.
{\tech} resolves calls and objects, constructs an AETG,
and processes events along multiple execution paths.
These operations add analysis work beyond the baseline checks.
The component measurements below identify where
the additional time and memory costs arise.

\begin{table}[htp]
\centering
\caption{Component and end-to-end analysis cost.}
\vspace{-10pt}
\label{tab:rq2-component-cost}
\smallskip
\scriptsize
\setlength{\tabcolsep}{3.8pt}
\renewcommand{\arraystretch}{1.10}

\resizebox{0.75\linewidth}{!}{%
\begin{tabular}{@{}lrrrrrr@{}}
\toprule
Component
  & \multicolumn{3}{c}{Time (s)}
  & \multicolumn{3}{c}{Peak Python heap (MB)} \\
\cmidrule(lr){2-4}\cmidrule(lr){5-7}
  & Mean & P95 & Max & Mean & P95 & Max \\
\midrule
Bytecode front end
  & 7.93 & 32.82 & 579.19
  & 10.8 & 41.3 & 481.2 \\
AETG construction
  & 25.64 & 109.24 & 1,854.82
  & 40.9 & 164.2 & 2,190.1 \\
Event simulation/inference
  & 35.29 & 172.66 & 977.76
  & 7.1 & 30.3 & 207.3 \\
End-to-end
  & 67.17 & 293.97 & 2,385.34
  & 57.5 & 211.0 & 2,599.9 \\
\bottomrule
\end{tabular}%
}
\end{table}

\vspace{2pt}
\noindent
\blackcircleone{2} \emph{Component cost.}
Table~\ref{tab:rq2-component-cost} shows distinct time and
memory patterns across the analysis components.
Event simulation/inference has the highest mean and P95
analysis times, at 35.29~s and 172.66~s.
AETG construction follows at 25.64~s and 109.24~s.
The memory results show the opposite pattern.
AETG construction has the highest mean and P95 phase-local
heap usage, at 40.9~MB and 164.2~MB,
compared with 7.1~MB and 30.3~MB for simulation/inference.
AETG construction also has the highest maximum component
time, reaching 1,854.82~s.
%
These differences reflect the work performed by each stage.
AETG construction stores instruction nodes, execution edges,
async events, and object relations.
Simulation traverses the constructed graph along selected
paths and updates event traces and abstract states.
The measurements are consistent with graph storage contributing
more heap usage and repeated path processing contributing
more analysis time.
The large maximum construction time also shows that
graph construction can be expensive for individual artifacts.

\begin{figure*}[htp]
\centering
\includegraphics[width=0.92\textwidth]
  {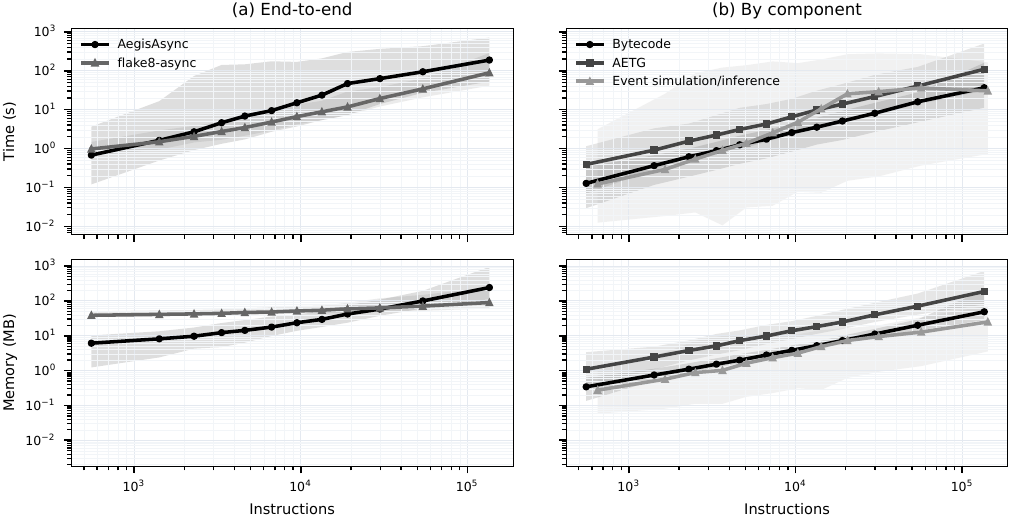}
\vspace{-10pt}
\caption{End-to-end (left) and component-level (right) cost
versus package instruction count.}
\label{fig:rq2-total-instruction-scaling}
\Description{End-to-end and component-level analysis time and memory
across instruction-count bins containing equal numbers of artifacts.
Curves show medians, and shaded regions show central 90 percent ranges.
Both tools show increasing costs with package size.
AegisAsync has lower median recorded peak memory in smaller bins
and higher median recorded peak memory in the largest bins.}
\end{figure*}

\vspace{2pt}
\noindent
\blackcircleone{3} \emph{Scaling with package size.}
We measure package size by compiled bytecode instruction count.
Figure~\ref{fig:rq2-total-instruction-scaling} groups artifacts
into bins containing equal numbers of artifacts.
In the left column, median analysis time for {\tech}
increases from 0.64~s in the smallest bin to 170.34~s
in the largest.
Both tools take longer as package size increases,
with {\tech} showing higher median times in larger bins.
The central 90\% ranges also widen, indicating greater
variation in analysis time among larger artifacts.
The memory comparison changes with package size.
{\tech} has lower median recorded peak memory in smaller bins,
but higher median memory in the largest bins.
These differences are not visible from the similar
overall means alone.
For {\tech}, instruction count is strongly associated
with both analysis time and peak Python heap usage.
Spearman $\rho$ is 0.807 for time and 0.936 for heap,
showing a stronger association with memory.
The component profiles in the right column show that
AETG construction has the highest median phase-local heap
across the displayed bins.
Its median analysis time also increases with instruction count,
while simulation/inference exhibits wider time ranges.
These patterns reflect the different scaling factors
in the analysis.
Larger packages contain more instructions and relationships
represented in the AETG, increasing the graph information stored.
Simulation cost also depends on the number of selected paths
and the events processed along them.
The component profiles are consistent with the stronger
association between instruction count and heap usage,
while variation in path processing contributes to
differences in analysis time among similarly sized packages.

\find{{\tech} completes analysis on all 2,500 sampled artifacts,
taking 67.17~s per artifact on average, or 3.3$\times$
the baseline's mean analysis time.
The path budget is reached in 7.5\% of analyses.
Simulation/inference has the highest mean component time,
while AETG construction has the highest phase-local heap usage.
Analysis cost increases with package size, with instruction
count more strongly associated with memory than time.}

\subsection{RQ3: Sensitivity}
\label{sec:eval:sensitivity}

RQ3 examines how the path budget affects finding yield and analysis
cost, and how the number of detector plugins affects inference cost.

\vspace{3pt}
\noindent
\textbf{\blackcircleone{1} Path-Budget Sensitivity.}
We vary $N_p\in\{1024,2048,4096,8192,16384\}$ while keeping
the AETG, detector set, and environment fixed.
The default is $N_p=4096$.
Table~\ref{tab:rq3-path-sensitivity} shows that reported candidates
increase from 380 to 910 and confirmed records from 184 to 457
as $N_p$ rises from 1,024 to 16,384.
The confirmed count continues to grow throughout the tested range,
with 118 additional records between the two largest budgets.
Among analyses below the configured bound, the mean path count
increases from 267 to 2,626, while the fraction reaching the
bound falls from 11.2\% to 5.0\%.
Larger budgets allow the simulator to explore additional paths,
exposing event sequences that the smaller budgets do not reach.
These paths produce both confirmed and unconfirmed candidates.
Record-level precision ranges from 44.7\% to 50.2\%
without increasing monotonically.
The results show that broader path exploration increases
confirmed finding yield, although some analyses still reach
the largest tested budget.

\begin{table}[htp]
\centering
\caption{Path-budget sensitivity results.}
\vspace{-10pt}
\label{tab:rq3-path-sensitivity}
\smallskip
\footnotesize
\setlength{\tabcolsep}{3.2pt}
\renewcommand{\arraystretch}{1.05}

\resizebox{0.6\columnwidth}{!}{%
\begin{tabular}{@{}lrrrrrr@{}}
\toprule
$N_p$ & Cand. & Conf. rec. & Prec.
& \multicolumn{2}{c}{Paths below bound}
& \multirow{2}{*}{At bound} \\
\cmidrule(lr){5-6}
& & & & Mean & P95 & \\
\midrule
1,024  & 380 & 184 & 48.4\% & 267   & 708    & 280 (11.2\%) \\
2,048  & 494 & 221 & 44.7\% & 471   & 1,459  & 227 (9.1\%) \\

\rowcolor{black!7}
4,096  & 572 & 259 & 45.3\% & 835   & 2,940  & 188 (7.5\%) \\

8,192  & 684 & 339 & 49.6\% & 1,492 & 5,927  & 152 (6.1\%) \\
16,384 & 910 & 457 & 50.2\% & 2,626 & 11,465 & 124 (5.0\%) \\
\bottomrule
\end{tabular}%
}

\vspace{2pt}
\parbox{\columnwidth}{\scriptsize
\textit{Note.} Mean and P95 are computed from runs below the
bound and rounded to the nearest path.
The shaded row marks the default budget.}
\end{table}

\begin{figure*}[htp]
\centering
\includegraphics[width=0.86\textwidth]
  {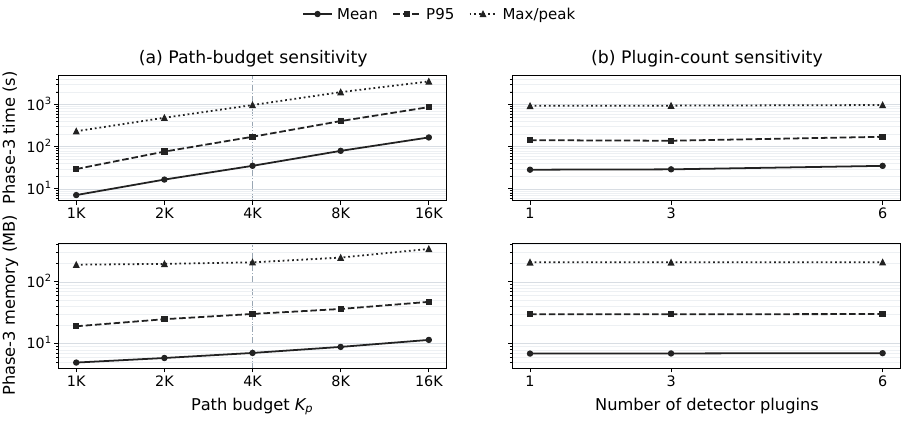}
\vspace{-10pt}
\caption{Cost sensitivity to path budget (left) and detector count
(right). The left column profiles simulation and inference
(Stages~2 and~3); the right profiles inference on precomputed
traces. Rows report analysis time and phase-local peak Python
heap usage. Curves show the mean, P95, and maximum.
The vertical marker in the left column denotes the default
$N_p=4096$.}
\label{fig:rq3-cost-sensitivity}
\end{figure*}

The left column of Figure~\ref{fig:rq3-cost-sensitivity} shows
that mean, P95, and maximum simulation/inference times increase
with $N_p$, while peak Python heap usage grows more gradually.
Larger budgets therefore expose additional findings at a higher
analysis cost, particularly in time spent processing paths.

\vspace{3pt}
\noindent
\textbf{\blackcircleone{2} Detector-Set Sensitivity.}
We evaluate 1, 3, and 6 detector plugins at the default
$N_p=4096$ over the same 2,500 artifacts.
The 1- and 3-plugin configurations use the first one and three
plugins listed in Appendix~\ref{ss:appplugins}.
All configurations reuse the same AETG and precomputed
trace/state pairs, isolating the cost of detector inference.
The right column of Figure~\ref{fig:rq3-cost-sensitivity} shows
that mean inference time increases from 28.6~s with one plugin
to 35.2~s with six, a 23.1\% increase.
Mean peak Python heap usage remains within 6.9--7.0~MB.
The additional plugins evaluate their risk predicates and
construct witnesses from the existing trace/state pairs.
They therefore reuse the results of graph construction and
path simulation rather than repeating these analyses.
The measurements show that the evaluated detector sets
add modest inference time with nearly unchanged heap usage.

\begin{figure*}[htp]
\centering
\includegraphics[width=0.92\textwidth]
  {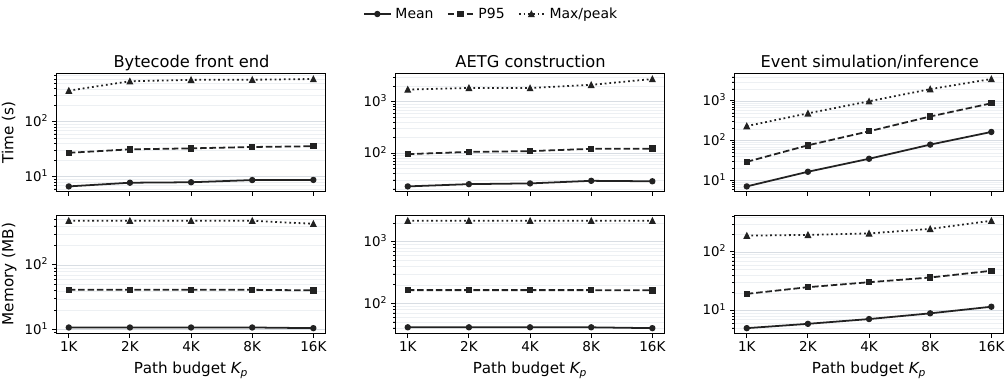}
\vspace{-10pt}
\caption{Component cost across path budgets. Each column shows
one analysis component; rows report analysis time and
phase-local peak Python heap usage.}
\label{fig:rq3-component-cost-curves}
\end{figure*}

\vspace{3pt}
\noindent
\textbf{\blackcircleone{3} Component-Level Path-Budget Sensitivity.}
Figure~\ref{fig:rq3-component-cost-curves} separates the
effect of increasing $N_p$ across the analysis components.
Bytecode processing and AETG construction remain nearly stable,
while simulation/inference accounts for most of the growth
in analysis time and phase-local peak Python heap.
The increase is more pronounced for time than memory.
This pattern follows the separation between graph construction
and path simulation.
The AETG is constructed once for each artifact, while $N_p$
controls how many paths are subsequently explored.
Larger budgets add event processing, state updates, and
detector evaluation without repeating graph construction.
The component results show that the cost of broader exploration
is concentrated in path-local event simulation and risk inference.

\vspace{3pt}
\find{Increasing the path budget from 1,024 to 16,384 raises
confirmed records from 184 to 457, with additional cost
concentrated in simulation and inference.
Bytecode processing and AETG construction remain nearly stable
across budgets.
On shared traces, expanding from one detector plugin to six
increases mean inference time by 23.1\%, while mean peak
Python heap usage remains nearly unchanged.}

\subsection{RQ4: Practical Utility}
\label{sec:eval:package}

RQ4 evaluates the candidates reported in real-world packages
through source and witness review.
We examine the candidate workload, confirmed bugs, and
the evidence used to assess reported risks in their
application context.

\vspace{3pt}
\noindent
\blackcircleone{1} \emph{Review protocol.}
We use a three-person review protocol
~\cite{ralph2021acm,votipka2020understanding}.
Two reviewers independently classify each record as
confirmed or unconfirmed using a predefined rubric.
Confirmation requires a defect supported by the source code
and an execution path exhibiting the reported behavior.
For {\tech}, reviewers also check whether the witness
matches the event order and object relationships in the source.
A third reviewer resolves disagreements, and ambiguous
records are classified as unconfirmed.
Both tools use the same source-level confirmation criteria.
We review all {\tech} candidates after deduplicating reports
across the five path budgets evaluated in RQ3.
For \textsc{flake8-async}, we sample diagnostics separately
within each shared risk family.
The finite-population sample-size formula~\cite{cochran1977sampling},
using 95\% confidence, a 5\% margin of error, and $p=0.5$,
gives 245 Task lifetime, 128 Blocking operation, and
356 Cancellation suppression diagnostics per round.
Three fixed-seed rounds produce 729 review judgments each,
or 2,187 in total.
Sites may appear in multiple rounds, and sampling fractions
differ across families.
The pooled confirmation rate therefore describes the
reviewed judgments.

\vspace{3pt}
\noindent
\blackcircleone{2} \emph{Candidate workload and confirmation.}
Table~\ref{tab:rq4-candidate-workload} shows that {\tech}
reports 484 candidates in the three shared families,
compared with 5,581 \textsc{flake8-async} diagnostic sites.
The difference is concentrated in Task lifetime and
Cancellation suppression, where {\tech} reports 64 and
175 candidates compared with 674 and 4,717 baseline diagnostics.
For Blocking operation, {\tech} reports more candidates:
245 compared with 190.
The detectors evaluate operations using recovered object
relationships and execution histories.
Task-lifetime checks examine task observation and ownership
at scope exit, while cancellation checks connect delivery
to handler outcomes along simulated paths.
Blocking checks use resolved call targets and their async
execution context.
These mechanisms provide the event and state evidence
used to assess each reported condition.
Source review confirms 457 of the 910 {\tech} candidates
(50.2\%), representing 457 unique bugs
(Table~\ref{tab:rq4-review-overall}).
Within the shared families, 201 of 484 candidates are
confirmed (41.5\%).
The baseline review confirms 70 of 2,187 judgments (3.2\%),
representing 33 unique sites.
Matching reports by package, family, file, and witness context
shows that {\tech} reports all 33 confirmed baseline sites.
The smaller shared-family candidate set therefore retains
the confirmed sites identified by sampled baseline review.

\begin{table}[htp]
\centering
\begin{minipage}[t]{0.47\linewidth}
\vspace{0pt}
\centering
\caption{Candidate workload by risk family.}
\vspace{-10pt}
\label{tab:rq4-candidate-workload}
\smallskip
\scriptsize
\setlength{\tabcolsep}{2.5pt}
\renewcommand{\arraystretch}{1.08}

\begin{tabularx}{\linewidth}{@{}
>{\raggedright\arraybackslash}Xrr@{}}
\toprule
Family & {\tech} & \textsc{flake8-async} \\
\midrule
Task lifetime            & 64  & 674 \\
Blocking operation       & 245 & 190 \\
Admission/fan-out        & 379 & -- \\
Cancellation suppression & 175 & 4,717 \\
Queue admission          & 32  & -- \\
Missing timeout          & 15  & -- \\
\midrule
Shared scope             & 484 & 5,581 \\
All {\tech} families     & 910 & -- \\
\bottomrule
\end{tabularx}
\end{minipage}
\hfill
\begin{minipage}[t]{0.50\linewidth}
\vspace{0pt}
\centering
\caption{Overall source-review results.}
\vspace{-10pt}
\label{tab:rq4-review-overall}
\smallskip
\scriptsize
\setlength{\tabcolsep}{2.5pt}
\renewcommand{\arraystretch}{1.08}

\begin{tabularx}{\linewidth}{@{}
>{\raggedright\arraybackslash}Xrrrr@{}}
\toprule
Tool/sample & Reviewed & Conf. & Unique & Rate \\
\midrule
{\tech} (all $N_p$)
  & 910 & 457 & 457 & 50.2\% \\
\midrule
\textsc{flake8-async}, round 1
  & 729 & 23 & -- & 3.2\% \\
\textsc{flake8-async}, round 2
  & 729 & 20 & -- & 2.7\% \\
\textsc{flake8-async}, round 3
  & 729 & 27 & -- & 3.7\% \\
\midrule
\textsc{flake8-async}, pooled
  & 2,187 & 70 & 33 & 3.2\% \\
\bottomrule
\end{tabularx}

\smallskip
\raggedright
Baseline counts are review judgments.
``Unique'' counts confirmed bugs for {\tech}
and confirmed sites for the baseline.
\end{minipage}
\end{table}

\vspace{3pt}
\noindent
\blackcircleone{3} \emph{Differences across risk families.}
Table~\ref{tab:rq4-family-validation} shows that confirmed
bugs are concentrated in Blocking operation and Admission/fan-out.
These families contribute 176 and 241 bugs, with confirmation
rates of 71.8\% and 63.6\%, respectively.
Together, they account for 417 of the 457 confirmed bugs (91.2\%).
These findings exercise two capabilities of the analysis.
For Blocking operation, call resolution identifies synchronous
operations, while the trace records their async execution context.
For Admission/fan-out, the analysis connects admitted work to
task-creation operations and available admission bounds.
Source review confirms that these facts support the reported
defect mechanisms in the analyzed packages.
The high confirmed yield in these families shows the practical
value of recovering call effects and admission behavior.

\begin{table}[t]
\centering
\caption{Source-review results by detector family.}
\vspace{-10pt}
\label{tab:rq4-family-validation}
\smallskip
\scriptsize
\setlength{\tabcolsep}{4pt}
\renewcommand{\arraystretch}{1.08}

\begin{tabularx}{\linewidth}{@{}
>{\raggedright\arraybackslash}Xrrrrrrrr@{}}
\toprule
& \multicolumn{4}{c}{{\tech}}
& \multicolumn{4}{c}{\textsc{flake8-async}} \\
\cmidrule(lr){2-5}\cmidrule(lr){6-9}
Family
& Reviewed & Conf. & Unique & Rate
& Reviewed & Conf. & Unique & Rate \\
\midrule
Task lifetime
& 64 & 9 & 9 & 14.1\%
& 735 & 5 & 4 & 0.7\% \\
Blocking operation
& 245 & 176 & 176 & 71.8\%
& 384 & 65 & 29 & 16.9\% \\
Admission/fan-out
& 379 & 241 & 241 & 63.6\%
& -- & -- & -- & -- \\
Cancellation suppression
& 175 & 16 & 16 & 9.1\%
& 1,068 & 0 & 0 & 0.0\% \\
Queue admission
& 32 & 7 & 7 & 21.9\%
& -- & -- & -- & -- \\
Missing timeout
& 15 & 8 & 8 & 53.3\%
& -- & -- & -- & -- \\
\midrule
Shared scope
& 484 & 201 & 201 & 41.5\%
& 2,187 & 70 & 33 & 3.2\% \\
All {\tech} families
& 910 & 457 & 457 & 50.2\%
& -- & -- & -- & -- \\
\bottomrule
\end{tabularx}
\end{table}

Confirmation rates are lower for Task lifetime and
Cancellation suppression, at 9 of 64 candidates (14.1\%)
and 16 of 175 (9.1\%), respectively.
Their witnesses record task observation and ownership
through scope exit, or cancellation delivery and handler outcomes.
Reviewers examine these histories together with the application's
task ownership and cancellation behavior.
For Cancellation suppression, the observed rate corresponds
to about 11 candidate reviews per confirmed bug.
These families therefore require more candidate reviews
per confirmed finding.
The three families outside the baseline comparison,
Admission/fan-out, Queue admission, and Missing timeout,
contribute 256 confirmed bugs.
Queue admission confirms 7 of 32 candidates (21.9\%),
while Missing timeout confirms 8 of 15 (53.3\%).
These detectors evaluate different admission and timeout
conditions using the same AETG and simulated traces.
Confirmed findings across all six families show that
the shared representation supports multiple async-risk
predicates.

\vspace{3pt}
\noindent
\blackcircleone{4} \emph{Distribution across interpreter versions.}
Table~\ref{tab:bugs-family-version} groups the 457 unique
source-confirmed bugs by the CPython version assigned
to each artifact.
Each interpreter group contains 500 sampled artifacts,
with different packages in each group.
Confirmed bugs occur in all five groups.
The 3.12 group contributes the largest share, with
179 bugs (39.2\%), including 126 Blocking operation findings.
Admission/fan-out accounts for most findings in the
3.11 and 3.13 groups, with 88 and 70 bugs, respectively.
These differences describe the composition of the sampled
packages under each assigned interpreter.
The version-aware front end produces bytecode representations
used by the same AETG construction, simulation, and detector
pipeline.
The confirmed findings show that
the analysis produces useful defect evidence under each
supported CPython version.

\begin{table}[htp]
\centering
\caption{Unique source-confirmed bugs by detector family
and assigned CPython version.}
\vspace{-10pt}
\label{tab:bugs-family-version}
\smallskip
\scriptsize
\setlength{\tabcolsep}{5pt}
\renewcommand{\arraystretch}{1.08}

\begin{tabularx}{\linewidth}{@{}
>{\raggedright\arraybackslash}Xrrrrrr@{}}
\toprule
Family & 3.10 & 3.11 & 3.12 & 3.13 & 3.14 & Total \\
\midrule
Task lifetime
  & 2 & 1 & 3 & 1 & 2 & 9 \\
Blocking operation
  & 21 & 16 & 126 & 8 & 5 & 176 \\
Admission/fan-out
  & 21 & 88 & 45 & 70 & 17 & 241 \\
Cancellation suppression
  & 8 & 1 & 4 & 0 & 3 & 16 \\
Queue admission
  & 2 & 3 & 0 & 1 & 1 & 7 \\
Missing timeout
  & 0 & 4 & 1 & 0 & 3 & 8 \\
\midrule
Total
  & 54 (11.8\%) & 113 (24.7\%) & 179 (39.2\%)
  & 80 (17.5\%) & 31 (6.8\%) & 457 (100\%) \\
\bottomrule
\end{tabularx}
\end{table}

\find{Source review confirms 457 of 910 {\tech} candidates
(50.2\%) across all six families and five interpreter groups.
In the shared families, {\tech} reports 484 candidates compared
with 5,581 baseline diagnostic sites and includes all 33
confirmed baseline sites found by sampled review.
Blocking operation and Admission/fan-out contribute 417
confirmed bugs, while Task lifetime and Cancellation suppression
have lower confirmation yields.}

\subsection{Bug Validation}

We validate source-confirmed findings at runtime using
proof-of-concept (PoC) harnesses.
Findings with the same package release, risk family, and
root cause form one PoC target.
We count one PoC per target whose harness reproduces
the reported runtime symptom.
The 457 source-confirmed bugs span 142 package releases.
Table~\ref{tab:bug-poc-status} reports 166 reproduced PoCs
covering all six risk families.
Admission/fan-out accounts for 113 PoCs, followed by
Blocking operation with 17 and Cancellation suppression
with 16.
The remaining 20 cover Task lifetime, Queue admission,
and Missing timeout.
These reproductions connect the source-confirmed findings
to symptoms observed during execution.
By submission, we had reported 157 bugs to developers,
of which 16 had been confirmed and fixed.
The \artifact{artifact} includes all source-confirmed findings
and reproduced PoCs, together with developer issue records
and their confirmation and fix status.

\begin{table}[htp]
\centering
\caption{Runtime reproduction and developer fixes for source-confirmed
bugs.}
\vspace{-10pt}
\label{tab:bug-poc-status}
\smallskip
\scriptsize
\setlength{\tabcolsep}{3.2pt}
\renewcommand{\arraystretch}{1.0}

\resizebox{0.65\linewidth}{!}{%
\begin{tabular}{@{}lrrrr@{}}
\toprule
Family & Unique bugs & Releases & PoCs
       & \shortstack{Confirmed and fixed} \\
\midrule
Task lifetime
  & 9   & 7   & 7   & 1 \\
Blocking operation
  & 176 & 34  & 17  & 0 \\
Admission/fan-out
  & 241 & 96  & 113 & 10 \\
Cancellation suppression
  & 16  & 11  & 16  & 3 \\
Queue admission
  & 7   & 6   & 6   & 0 \\
Missing timeout
  & 8   & 7   & 7   & 2 \\
\midrule
Overall
  & 457 & 142 & 166 & 16 \\
\bottomrule
\end{tabular}%
}
\end{table}

\vspace{3pt}
\noindent
\textbf{Case studies.}
The following cases cover \emph{Blocking operation},
\emph{Admission/fan-out}, and \emph{Cancellation suppression}.
Each case explains the code pattern, the witness produced by
{\tech}, and the runtime symptom reproduced by its PoC.

\vspace{3pt}
\noindent
\textbf{Case 1: Blocking operation.}

\begin{tcolorbox}[
  colback=black!3,colframe=black!30,boxrule=0.3pt,arc=0.5mm,
  width=0.97\linewidth,
  left=0.8mm,right=0.8mm,top=0.5mm,bottom=0.5mm]
\scriptsize
\textbf{Package:} \texttt{spectryn} \hfill \textbf{CPython:} 3.11\\
\textbf{Location:} \texttt{src/spectryn/cli/commands/rest\_api.py:190}\\
\textbf{Function:} \texttt{run\_rest\_api\_async}
\end{tcolorbox}

\noindent{\scriptsize\textit{Async caller (lines 188--190):}}
\begin{lstlisting}[style=aegisasync-python,basicstyle=\ttfamily\scriptsize,
  numbers=left,numberstyle=\tiny,firstnumber=188,breaklines=true,
  columns=fullflexible]
# This implementation reuses the sync version for simplicity
# A full async implementation could use aiohttp or similar
return run_rest_api(args)
\end{lstlisting}

\noindent{\scriptsize\textit{Synchronous callee (lines 111--124):}}
\begin{lstlisting}[style=aegisasync-python,basicstyle=\ttfamily\scriptsize,
  numbers=left,numberstyle=\tiny,firstnumber=111,breaklines=true,
  columns=fullflexible]
# Run server (synchronous start, then wait for shutdown)
server._start_sync()

# Keep running until shutdown
try:
    while server.is_running() and not shutdown_event.is_set():
        import time

        time.sleep(0.5)
except KeyboardInterrupt:
    pass

# Stop server
server._stop_sync()
\end{lstlisting}
The async entry point \texttt{run\_rest\_api\_async} directly calls
the synchronous \texttt{run\_rest\_api} function.
The callee starts the server and waits for shutdown in a loop
that repeatedly calls \texttt{time.sleep(0.5)}.
When the async entry point runs on the event-loop thread,
the synchronous call blocks other coroutines until the server
shuts down.
This delays request processing, timeouts, and cancellation
handling, potentially making the service unresponsive under
concurrent traffic and causing denial of service.
The AETG connects the async entry point to the synchronous
callee, and simulation records the
\texttt{sync\_call\_on\_async\_path} fact.
The Blocking operation detector uses this evidence to report
line~190.
The PoC keeps the async wrapper active for approximately 1.5~s,
during which all 30 expected heartbeat opportunities are missed,
demonstrating event-loop starvation.


\vspace{3pt}
\noindent
\textbf{Case 2: Admission/fan-out.}

\begin{tcolorbox}[
  colback=black!3,colframe=black!30,boxrule=0.3pt,arc=0.5mm,
  width=0.97\linewidth,
  left=0.8mm,right=0.8mm,top=0.5mm,bottom=0.5mm]
\scriptsize
\textbf{Package:} \texttt{clinicaltrials-mcp} \hfill \textbf{CPython:} 3.12\\
\textbf{Location:} \texttt{src/clinicaltrials\_mcp/client.py:165--166}\\
\textbf{Function:} \texttt{ClinicalTrialsClient.get\_studies}
\end{tcolorbox}

\noindent{\scriptsize\textit{Admission site (lines 165--167):}}
\begin{lstlisting}[style=aegisasync-python,basicstyle=\ttfamily\scriptsize,
  numbers=left,numberstyle=\tiny,firstnumber=165,breaklines=true,
  columns=fullflexible]
tasks = [self.get_study(nct_id) for nct_id in nct_ids]
results = list(await asyncio.gather(*tasks))
return results
\end{lstlisting}

\noindent{\scriptsize\textit{Per-item HTTP operation (lines 140--141):}}
\begin{lstlisting}[style=aegisasync-python,basicstyle=\ttfamily\scriptsize,
  numbers=left,numberstyle=\tiny,firstnumber=140,breaklines=true,
  columns=fullflexible]
nct_id = nct_id.strip().upper()
response = await self._client().get(f"/studies/{nct_id}")
\end{lstlisting}
The function schedules all requested study operations without
limiting concurrent admission.
The comprehension creates one coroutine per identifier in
\texttt{nct\_ids}, and \texttt{asyncio.gather} schedules the
entire collection.
A caller triggers the issue by supplying a large collection,
causing many coroutines and HTTP operations to be scheduled.
This can increase memory consumption, delay other requests,
and place excessive load on the client connection pool and
downstream service, potentially causing resource-exhaustion
denial of service.
An HTTP connection pool may limit simultaneous connections
but does not bound the number of scheduled coroutines.
The AETG relation \texttt{fanout\_site} identifies the
input-dependent admission point.
Simulation records the event chain
\texttt{Fanout} $\rightarrow$
\texttt{CallUse(asyncio.gather)} $\rightarrow$
\texttt{FanoutAdmission}.
With no finite \texttt{FanoutBound} fact, the Admission/fan-out
detector reports line~166.
The PoC supplies 512 identifiers and uses a controlled
per-item operation.
All 512 coroutines are admitted before any is released,
confirming that admission grows with the input size.

\vspace{3pt}
\noindent
\textbf{Case 3: Cancellation suppression.}

\begin{tcolorbox}[
  colback=black!3,colframe=black!30,boxrule=0.3pt,arc=0.5mm,
  width=0.97\linewidth,
  left=0.8mm,right=0.8mm,top=0.5mm,bottom=0.5mm]
\scriptsize
\textbf{Package:} \texttt{apexnova} \hfill \textbf{CPython:} 3.10\\
\textbf{Location:} \texttt{src/apexnova/stub/resilience/}\\
\hspace*{12mm}\texttt{resilience\_patterns.py:746--755}\\
\textbf{Function:} \texttt{ResiliencePolicy.execute}
\end{tcolorbox}

\noindent{\scriptsize\textit{Primary operation path (lines 730--740):}}
\begin{lstlisting}[style=aegisasync-python,basicstyle=\ttfamily\scriptsize,
  numbers=left,numberstyle=\tiny,firstnumber=730,breaklines=true,
  columns=fullflexible]
try:
    if self.retry:
        result = await self.retry.execute(wrapped_operation)
    else:
        result = await wrapped_operation()

    execution_time = timedelta(seconds=time.time() - start_time)
    return Success(value=result, execution_time=execution_time)

except Exception as e:
\end{lstlisting}

\noindent{\scriptsize\textit{Fallback cancellation path (lines 743--755):}}
\begin{lstlisting}[style=aegisasync-python,basicstyle=\ttfamily\scriptsize,
  numbers=left,numberstyle=\tiny,firstnumber=743,breaklines=true,
  columns=fullflexible]
# Try fallback
if self.fallback:
    try:
        fallback_result = await self.fallback.execute(lambda: None)
        return Fallback(
            value=fallback_result,
            original_error=e,
            execution_time=execution_time,
        )
    except:
        pass

return Failure(error=e, execution_time=execution_time)
\end{lstlisting}
The fallback handler uses a bare \texttt{except} that catches
\texttt{CancelledError} without propagating it.
When the primary operation fails, the policy enters its async
fallback.
If the policy task is cancelled while awaiting the fallback
at line~746, cancellation reaches the bare handler at line~752.
The handler discards the exception and returns an ordinary
\texttt{Failure} result.
This suppresses the cancellation signal, causing the caller
to treat the cancelled operation as a normal failure.
In applications that rely on cancellation to enforce deadlines
or stop pending work, this can undermine timeout and shutdown
controls, allowing further processing or delaying termination.
Path simulation connects cancellation at the fallback await
to the bare handler and its subsequent normal return.
The Cancellation suppression detector uses this trace to
report line~752.
The PoC forces the primary operation to fail and cancels
the policy task during the fallback await.
The task returns \texttt{Failure} instead of raising
\texttt{CancelledError}, confirming cancellation suppression.

\section{Discussion}\label{sec:diss}

\noindent
\blackcircleone{1} \emph{Preserving detector inputs across versions.}
Version adapters translate CPython instructions and exception
metadata into common instruction categories.
AETG construction and simulation turn these instructions
into the events, object relationships, and states used by
the detectors.
Updating an adapter therefore requires checking that these
detector inputs are preserved.
The evidence checks in Section~\ref{sec:eval:effectiveness}
provide regression criteria for changes to version adapters
and operation-recognition rules.
Coverage can be extended by adding cases that specify the
events, bindings, and execution order expected from new
bytecode patterns or recognized calls.

\vspace{3pt}
\noindent
\blackcircleone{2} \emph{Exploration and detector cost.}
The implementation separates path exploration from risk inference.
The path budget bounds initial path selection, and the
static-edge restriction bounds repeated traversal within a path.
Section~\ref{sec:eval:sensitivity} shows that larger budgets
expose additional findings, with cost growth concentrated
in simulation and inference.
Detectors whose conditions use existing event and state facts
can reuse the constructed AETG and simulated traces.
The detector-count experiment measures the 
inference for such extensions.
A risk condition that requires new event kinds or state
properties requires changes to event generation or
state-transfer functions.
The shared interface therefore supports detector reuse
while making the required analysis inputs explicit.

\vspace{3pt}
\noindent
\blackcircleone{3} \emph{Using witnesses to refine the analysis.}
Each candidate links a predicate match to the event order,
symbolic object bindings, and resolution facts supporting it.
These records help distinguish errors in the recovered
behavior from errors in the risk condition applied to it.
For example, reviewing a cancellation witness involves
checking the exception path and then assessing the handler's
response in the application's cancellation protocol.
Recording why candidates remain unconfirmed would guide
changes to operation recognition, points-to precision,
path conditions, or detector predicates.
The lower confirmation rates for Task lifetime and
Cancellation suppression in Section~\ref{sec:eval:package}
make their reports priorities for this refinement.

\section{Related Work}\label{sec:related}

\noindent
\blackcircleone{1} \emph{Event-driven and asynchronous program analysis.}
Madsen et al.~\cite{nodejs_event_analysis} use event-based call graphs
to model listener registration and event emission in Node.js.
Their event- and listener-sensitive analyses distinguish dataflow
by emitted events and registered listeners.
Sotiropoulos and Livshits~\cite{async_js_analysis} use callback
graphs to guide dataflow analysis of promises, timers, and
asynchronous I/O according to callback execution order.
Yee et al.~\cite{yee2019precise} augment dataflow analysis with
event-handler states to filter infeasible interprocedural paths.
{\tech} connects Python task creation, scheduling, and observation
through symbolic objects in the AETG.
Simulation tracks task and resource states along selected paths,
allowing detectors to combine object relationships with event order.

\noindent
\blackcircleone{2} \emph{Cancellation analysis.}
Sethi et al.~\cite{cancellation_systems} study cancellation failures
in Java, C\#, and Go systems and develop CodeQL checkers for
Java and C\#.
Their checks cover unhandled interrupt exceptions in loops,
uncancelled child tasks, and ignored cancellation tokens.
{\tech} explores Python cancellation from the trace and abstract
state reached at an await suspension.
Each continuation follows handlers and cleanup while carrying
forward earlier task and resource effects.
The trace links cancellation delivery to the handler occurrence
that receives it and records whether the handler propagates
cancellation or continues normally.

\noindent
\blackcircleone{3} \emph{General-purpose Python analysis.}
Bandit checks Python syntax trees for security issues~\cite{bandit},
Pyre checks types~\cite{pyre}, and PyCG derives call graphs from
interprocedural assignment relations~\cite{salis2021pycg}.
The PyFlow preprint~\cite{gu2026pyflow} shares an interprocedural
dataflow solver across taint, nullness, and typestate clients.
Its control flow represents exception and suspension boundaries.
The CPyGraph preprint~\cite{chen2026cpygraph} provides version-aware
points-to, call-graph, control-flow, and dependence analysis of
CPython bytecode, including coroutine instructions.
Its graph products share bytecode identities and resolution facts.
{\tech} combines call and object resolution with async-event
generation and simulation.
Its predicates evaluate task observation and ownership, resource
liveness at awaits, admission bounds, and cancellation outcomes
over the resulting traces and abstract states.

\noindent
\blackcircleone{4} \emph{Python async checkers.}
\textsc{flake8-async}~\cite{flake8-async} checks blocking calls,
task references, cancellation propagation, and checkpoints
across branches and loops.
Its rules flag discarded \texttt{asyncio.create\_task} results
and cancellation-catching handlers that return or raise another
exception.
{\tech} uses resolved calls and object flow to connect operations
across analyzed function bodies.
A task returned by a helper can be related to later uses in
the caller.
Task predicates use these relationships to evaluate observation
and ownership at creator exit.
Call resolution also exposes blocking operations within analyzed
callees.
Each candidate retains the relevant operations, their order,
and supporting state facts as a witness.

\noindent
\blackcircleone{5} \emph{Abstract interpretation and symbolic execution.}
Abstract interpretation approximates program behavior using
abstract domains and transfer functions~\cite{cousot1977abstract}.
KLEE~\cite{cadar2008klee} explores program paths with symbolic
inputs and solves path constraints to generate tests.
{\tech} propagates finite properties of symbolic tasks and
resources along bounded paths and retains event histories.
These properties can record earlier behavior, such as a resource
being live at an await before its later release.
Detectors evaluate the resulting trace/state pairs and retain
the events and properties supporting each match.
\section{Conclusion}\label{sec:conclusion}

This paper presents {\tech}, a bytecode-based analysis
of security and robustness risks in Python async programs.
Its AETG combines control flow, async events, and
symbolic-object relations.
Path-local simulation tracks event order and task states,
including cancellation delivery and handler outcomes.
Six detectors evaluate the resulting traces and states
and report candidates with supporting witnesses.
On AsyncBench, {\tech} detects all 180 labeled risks without
false positives across CPython~3.10--3.14.
Across five path budgets on 2,500 package versions, source review
confirms 457 unique bugs from 910 deduplicated candidates,
and 166 executable PoCs reproduce the reported behaviors.
The findings span all six risk families, showing that shared graph
and trace evidence supports different async-risk checks and
connects reported bugs to their execution histories.


\newpage
\bibliographystyle{ACM-Reference-Format}
\bibliography{reference}

\appendixcontent
\section{Detector Plugin Specifications}
\label{ss:appplugins}

Section~\ref{sec:phase3:event-trace-risk-inference} defines the
plugin interface and presents the cancellation-suppression detector.
Here we specify the remaining five plugins using
$d_i=\langle id_i,Q_i,P_i,W_i\rangle$.
Each plugin evaluates a trace/state pair $(\pi,\sigma_\pi)$:
$Q_i$ selects anchors, $P_i$ checks the risk condition,
and $W_i$ constructs the candidate and its supporting witness.
We use $x_i$ to denote an event occurrence in $\pi$.
Object anchors identify symbolic tasks or queues, while event
anchors identify operations on the trace.
The predicate $\mathsf{Resolved}$ checks whether the call target
or object binding required by the detector is resolved.
The notation $\sigma_\pi\models f$ means that the abstract
state records fact $f$.

\vspace{3pt}
\noindent
\blackcircleone{1} \emph{Task lifetime} (\texttt{AEGISASYNC001}).
This detector reports tasks that may remain live at their
creator's scope exit without recorded observation or
ownership transfer:
\[
\begin{aligned}
Q_{\mathit{task}}(\pi,\sigma_\pi)
=\{\,t\mid {}&
\exists x_i,x_j\in\pi.\;
x_i.kind=\mathsf{TaskCreate}
\land x_i.obj=t\\
&{}\land x_j.kind=\mathsf{ScopeExit}
\land x_i\prec_\pi x_j\,\},\\[2pt]
P_{\mathit{task}}(t,\pi,\sigma_\pi)
\equiv {}&
\mathsf{Resolved}(t)
\land
\bigl(\sigma_\pi\models\mathsf{taskLiveAtExit}(t)\bigr)\\
&{}\land
\neg\bigl(\sigma_\pi\models\mathsf{taskObserved}(t)\bigr)\\
&{}\land
\neg\bigl(\sigma_\pi\models
\mathsf{taskOwnershipTransfer}(t)\bigr).
\end{aligned}
\]
The fact $\mathsf{taskLiveAtExit}(t)$ records possible liveness
at the creator's exit, with task completion unresolved.
The observation and ownership predicates concern the same
symbolic task on that trace.
$W_{\mathit{task}}$ retains the creation event, task identity,
creator exit, and any recorded spawn or escape relations.
The witness also includes observation and ownership information,
source and resolution facts, and unresolved dependencies.
Unobserved tasks can hide failures, and repeated creation
of tasks that remain live can exhaust resources.

\vspace{3pt}
\noindent
\blackcircleone{2} \emph{Blocking operation} (\texttt{AEGISASYNC002}).
This detector reports synchronous blocking operations on
async paths without recorded offloading:
\[
\begin{aligned}
Q_{\mathit{block}}(\pi,\sigma_\pi)
&=\{\,x_i\in\pi\mid
x_i.kind=\mathsf{BlockingInAsync}\,\},\\[2pt]
P_{\mathit{block}}(x_i,\pi,\sigma_\pi)
&\equiv
\mathsf{Resolved}(x_i)
\land
\bigl(\sigma_\pi\models\mathsf{blockingInAsync}(x_i)\bigr)\\
&\qquad{}\land
\neg\bigl(\sigma_\pi\models\mathsf{offloaded}(x_i)\bigr).
\end{aligned}
\]
The fact $\mathsf{blockingInAsync}(x_i)$ records a blocking
operation encountered during async execution.
For parameter-dependent behavior, this fact requires
argument bindings that establish the blocking effect.
The fact $\mathsf{offloaded}(x_i)$ records recognized execution
outside the event-loop thread.
$W_{\mathit{block}}$ retains the call node, resolved target
and effect, async execution context, and relevant argument bindings.
Blocking the event loop can delay requests, timeouts,
and cancellation handling.

\vspace{3pt}
\noindent
\blackcircleone{3} \emph{Admission/fan-out} (\texttt{AEGISASYNC004}).
This detector reports operations that admit concurrent work
without a recorded finite bound:
\[
\begin{aligned}
Q_{\mathit{fan}}(\pi,\sigma_\pi)
&=\{\,x_i\in\pi\mid
x_i.kind=\mathsf{FanoutAdmission}\,\},\\[2pt]
P_{\mathit{fan}}(x_i,\pi,\sigma_\pi)
&\equiv
\mathsf{Resolved}(x_i)
\land
\bigl(\sigma_\pi\models\mathsf{fanoutAdmission}(x_i)\bigr)\\
&\qquad{}\land
\neg\bigl(\sigma_\pi\models\mathsf{fanoutBound}(x_i)\bigr).
\end{aligned}
\]
The fact $\mathsf{fanoutAdmission}(x_i)$ records admission
of the associated work, and $\mathsf{fanoutBound}(x_i)$
records a finite limit for that admission.
$W_{\mathit{fan}}$ retains the operation node, admitted
collection or object, resolved admission relation, and
bound information.
Input-dependent fan-out can turn one request into many
concurrent operations and exhaust local or downstream resources.

\vspace{3pt}
\noindent
\blackcircleone{4} \emph{Queue admission} (\texttt{AEGISASYNC005}).
This detector reports admission to async queues without
a recorded finite positive capacity:
\[
\begin{aligned}
Q_{\mathit{queue}}(\pi,\sigma_\pi)
&=\{\,q\mid \exists x_i\in\pi.\;
x_i.kind=\mathsf{QueueAdmission}
\land x_i.obj=q\,\},\\[2pt]
P_{\mathit{queue}}(q,\pi,\sigma_\pi)
&\equiv
\mathsf{Resolved}(q)
\land
\bigl(\sigma_\pi\models\mathsf{queueAdmission}(q)\bigr)\\
&\qquad{}\land
\neg\bigl(\sigma_\pi\models\mathsf{queueBound}(q)\bigr).
\end{aligned}
\]
The fact $\mathsf{queueAdmission}(q)$ records an admission
operation involving queue $q$.
The fact $\mathsf{queueBound}(q)$ records a finite positive
capacity recovered from constructor arguments or propagated values.
$W_{\mathit{queue}}$ retains the queue identity, admission
occurrence, and capacity information.
An unbounded queue can accumulate objects when production
outpaces consumption, leading to memory exhaustion.

\vspace{3pt}
\noindent
\blackcircleone{5} \emph{Missing timeout} (\texttt{AEGISASYNC006}).
This detector reports resolved external async operations
followed by an await and a scope exit, with no recorded
cancellation-bound event on the trace:
\[
\begin{aligned}
Q_{\mathit{timeout}}(\pi,\sigma_\pi)
&=\{\,x_i\in\pi\mid
x_i.kind=\mathsf{ExternalAsyncOperation}\,\},\\[2pt]
P_{\mathit{timeout}}(x_i,\pi,\sigma_\pi)
&\equiv
\mathsf{Resolved}(x_i)\\
&\qquad{}\land
\bigl(
\exists x_j,x_k\in\pi.\;
x_i\prec_\pi x_j\prec_\pi x_k\\
&\qquad\qquad{}\land x_j.kind=\mathsf{Await}
\land x_k.kind=\mathsf{ScopeExit}
\bigr)\\
&\qquad{}\land
\neg\bigl(
\exists x_\ell\in\pi.\;
x_\ell.kind=\mathsf{CancellationBound}
\bigr).
\end{aligned}
\]
The ordered events identify the external operation,
a later await, and a scope exit.
A $\mathsf{CancellationBound}$ event records a recognized
timeout or cancellation bound.
$W_{\mathit{timeout}}$ retains the external operation,
await, scope context, and timeout or cancellation-bound information.
Pending operations can retain tasks and connections,
reducing the application's available capacity.
\section{Effectiveness by Case Structure on AsyncBench}
\label{app:rq1-structural}

Table~\ref{tab:asyncbench-structural} groups the cases by structure.
Single-risk cases isolate one defect, while mixed-risk cases
combine multiple risk families.
Systematic-stress cases exercise aliasing, interprocedural flow,
and multiple risk-relevant events.
The table reports the total labeled risks and the risks in
the three families shared with \textsc{flake8-async}.

\begin{table}[htp]
\centering
\caption{AsyncBench detection results by case structure.}
\vspace{-10pt}
\label{tab:asyncbench-structural}
\smallskip
\scriptsize
\setlength{\tabcolsep}{4pt}
\renewcommand{\arraystretch}{1.08}

\resizebox{0.75\linewidth}{!}{%
\begin{tabular}{@{}lrrrrr@{}}
\toprule
Case structure & Cases & Risks & Shared risks
               & {\tech} & \textsc{flake8-async} \\
\midrule
Single-risk
  & 28 & 28  & 23 & 28/0/0  & 6/0/17 \\
Mixed-risk
  & 13 & 33  & 18 & 33/0/0  & 9/0/9 \\
Systematic stress
  & 23 & 119 & 49 & 119/0/0 & 25/0/24 \\
Safe/negative
  & 14 & 0   & 0  & 0/0/0   & -- \\
\midrule
Overall
  & 78 & 180 & 90 & 180/0/0 & 40/0/50 \\
\bottomrule
\end{tabular}%
}

\smallskip
\parbox{0.85\linewidth}{\scriptsize
Shared risks belong to the three families evaluated by both tools.
A dash indicates an unreported result.}
\end{table}

{\tech} detects all labeled risks across the three positive
case groups and produces no reports on the 14 safe controls.
The systematic-stress group contains 119 of the 180 risks.
Detection across these cases shows that the analysis handles
the tested combinations of aliases, interprocedural flow,
and related async operations.
The mixed-risk results also show that multiple risk families
can be evaluated within the same case.
Within the shared families, \textsc{flake8-async} detects
6 of 23 single-risk instances, 9 of 18 mixed-risk instances,
and 25 of 49 systematic-stress instances.
Its 50 misses occur across all three groups.
{\tech} detects all 90 shared risks, including these missed instances.
The structural results complement the family-level results in RQ1.
They show that the tested event and object relationships remain
available to detection across isolated defects, mixed-risk cases,
and more complex program structures.

\end{document}